\documentclass[12pt]{article}
\usepackage[onehalfspacing]{setspace}
\usepackage[english]{babel}
\usepackage[margin=0.8in]{geometry}
\usepackage{amsmath,amsfonts,amssymb,mathrsfs}
\usepackage{bbm}
\usepackage{graphicx,subcaption}
\usepackage{hyperref}
\usepackage[round]{natbib}
\usepackage[toc,page]{appendix}
\usepackage{enumerate}
\usepackage{xr}
\usepackage{amsthm}

\hypersetup{
    colorlinks=true, 
    linktoc=all,        
    linkcolor=blue,  
    citecolor=blue,
    urlcolor=blue
}

\renewcommand{\P}{\mathbb{P}}
\newcommand{\E}{\mathbb{E}}

\newcommand{\1}{\mathbbm{1}}

\newtheorem{definition}{Definition}
\newtheorem{theorem}{Theorem}

\newtheorem{lemma}{Lemma}
\newtheorem{proposition}{Proposition}
\newtheorem{assumption}{Assumption}
\newtheorem{remark}{Remark}
\newtheorem{corollary}{Corollary}

\usepackage{xpatch} 
\makeatletter
\AtBeginDocument{\xpatchcmd{\@thm}{\thm@headpunct{.}}{\thm@headpunct{}}{}{}}
\makeatother

\newcommand\indep{\protect\mathpalette{\protect\independenT}{\perp}}
\def\independenT#1#2{\mathrel{\rlap{$#1#2$}\mkern2mu{#1#2}}}

\makeatletter 
\def\th@plain{%
  \thm@notefont{}
  \itshape 
}
\def\th@definition{%
  \thm@notefont{}
  \normalfont 
}
\makeatother

\title{Causal Spillover Effects Using Instrumental Variables\thanks{I thank Matias Cattaneo, Cl{\'e}ment de Chaisemartin, Xinwei Ma, Kenichi Nagasawa, Olga Namen, Neslihan Sakarya, Dick Startz and Doug Steigerwald for valuable discussions and suggestions, and seminar participants at Stanford University, UCSB Applied Microeconomics Lunch, Northwestern University, UCLA, UC San Diego, University of Chicago, UC Davis, USC and University of Essex for helpful comments. I also thank the editor, Jane-Ling Wang, the associate editor and three anonymous referees for their detailed suggestions that greatly improved the paper.}}

\author{Gonzalo Vazquez-Bare\footnote{Department of Economics, University of California, Santa Barbara. \href{gvazquez@econ.ucsb.edu}{gvazquez@econ.ucsb.edu}.}}

\begin{document}
\maketitle

\vspace{1 cm}

\abstract{I set up a potential outcomes framework to analyze spillover effects using instrumental variables. I characterize the population compliance types in a setting in which spillovers can occur on both treatment take-up and outcomes, and provide conditions for identification of the marginal distribution of compliance types. I show that intention-to-treat (ITT) parameters aggregate multiple direct and spillover effects for different compliance types, and hence do not have a clear link to causally interpretable parameters. Moreover, rescaling ITT parameters by first-stage estimands generally recovers a weighted combination of average effects where the sum of weights is larger than one. I then analyze identification of causal direct and spillover effects under one-sided noncompliance, and show that causal effects can be estimated by 2SLS in this case. I illustrate the proposed methods using data from an experiment on social interactions and voting behavior. I also introduce an alternative assumption, \textit{independence of peers' types}, that identifies parameters of interest under two-sided noncompliance by restricting the amount of heterogeneity in average potential outcomes.
}

\vspace{1 cm}


\setcounter{page}{0}\thispagestyle{empty}

\newpage

\setlength{\parskip}{.3cm}


\section{Introduction}\label{sec:intro}

An accurate assessment of spillover effects is crucial to learn about the costs and benefits of treatments  and policies \citep{Athey-Imbens_2017_handbook,Abadie-Cattaneo_2018_Annurev}. Previous literature has shown that appropriately designed randomized controlled trials (RCTs) are a powerful tool to analyze spillovers \citep{Moffit_2001,Duflo-Saez_2003_QJE,Hudgens-Halloran_2008_JASA,Baird-etal_2018_ReStat,Vazquez-Bare_2017_framework}. RCTs, however, are often subject to imperfect compliance, as individuals assigned to treatment may refuse it and individuals not assigned to treatment may find alternative sources to receive it. In other cases, researchers may not have control on the treatment assignment, and instead may need to rely on quasi-experimental variation from a natural experiment \citep[see e.g.][]{Angrist-Krueger_2001_JEP,Titiunik_2020_Chapter}. In such cases, actual treatment receipt becomes endogenous, even if treatment assignment is randomized (or as-if-randomized). While previous literature has shown that instrumental variables (IVs) can address this type of endogeneity and identify local average treatment effects when spillovers are ruled out \citep{Angrist-Imbens-Rubin_1996_JASA}, little is known about what an IV can identify in the presence of spillovers. 

This paper provides a framework to study causal spillover effects using instrumental variables and offers three main contributions. First, I define causal direct and spillover effects in a setup with two-sided noncompliance. When treatment take-up is endogenous, spillover effects can occur in two stages: (i) treatment take-up, when a unit's instrument can affect its peers' treatment status, and (ii) outcomes, when a unit's treatment can affect its peers' responses. Focusing on the case in which spillovers occur within pairs (such as spouses or roommates), I propose a generalization of the monotonicity assumption \citep{Imbens-Angrist_1994_ECMA} that partitions the population into five compliance types: in addition to always-takers, compliers and never-takers, units may be social-interaction compliers, who receive the treatment when either themselves or their peer are assigned to it, and group compliers, who only receive the treatment when both themselves and their peer are assigned to it. Proposition \ref{prop:comp_distr} provides conditions for identification of the marginal distribution of compliance types, and shows that the joint distribution is generally not identified. 

Second, I analyze intention-to-treat (ITT) parameters and show that these estimands conflate multiple direct and indirect effects for different compliance types, and hence do not have a clear causal interpretation in general. Moreover, rescaling the ITT by the first-stage estimand, which would recover the average effect on compliers in the absence of spillovers, generally yields a weighted average of direct and spillover effects where the sum of weights exceeds one. 

Third, I show that, when noncompliance is one-sided, it is possible to identify the average direct effect on compliers and the average spillover effect on units with compliant peers. I provide a way to assess the external validity of these parameters and discuss testable implications of the identification assumptions. In addition, I show that direct and indirect local average effects can be jointly estimated by two-stage least squares (2SLS). This provides a straightforward way to estimate these effects in practice based on standard regression methods and to conduct inference that is robust to weak instruments using Fieller-Anderson-Rubin confidence intervals. The proposed methods are illustrated using data from an experiment on social interactions in the household and voter turnout.

Finally, I discuss two generalizations of my results. The first one considers the case in which the IV identification assumptions hold after conditioning on a set of observed covariates. The second one generalizes the results to arbitrary group sizes by introducing a novel assumption, \textit{independence of peers' types}, by which potential outcomes are independent of peers' compliance types conditional on own type. I show how this alternative assumption permits identification under two-sided noncompliance by limiting the amount of heterogeneity in the distribution of potential outcomes.

This paper contributes to the literature on causal inference under interference, which generally focuses on the case of two-stage experimental designs with perfect compliance \citep[see][for a recent review]{Halloran-Hudgens_2016_CER}. Existing studies analyzing imperfect compliance \citep{Sobel_2006_JASA,Kang-Imbens_2016_wp,Kang-Keele_2018_wp,Imai-Jiang-Malani_2020_JASA} consider identification, estimation and inference for specific experimental designs or by imposing specific restrictions on the spillovers structure. My findings add to this literature by introducing a novel set of estimands and identification conditions that are independent of the experimental design and that simultaneously allow for spillovers on outcomes, spillovers on treatment take-up and multiple compliance types in a superpopulation setting. In related work, \citet{DiTraglia-etal_2021} analyze identification and estimation of direct and spillover effects in randomized saturation designs under one-sided noncompliance. Their approach is complementary to mine, as it assumes a specific experimental design and rules out the presence of spillovers in treatment take-up but focuses on the case of large and unequally-sized groups.

This paper can also be linked to the literature on multiple instruments \citep{Imbens-Angrist_1994_ECMA,Mogstad-etal_NBER_2019} since the vectors of own and peers' instruments and treatments in spillover analysis can be rearranged as a multivalued instrument and a multivalued treatment. While most of the existing literature in this area considers the case of multiple instruments and a single binary treatment, my setting with unrestricted spillovers introduces both multiple instruments and multiple treatments (see Remark \ref{rmk:multi} for further discussion).

The rest of the article is organized as follows. Section \ref{sec:setup} describes the setup, introduces the notation and defines the causal parameters of interest. Section \ref{sec:ITT} analyzes ITT parameters. Section \ref{sec:identif} provides the main identification results under one-sided noncompliance. Section \ref{sec:estimation} analyzes estimation and inference, and Section \ref{sec:emp_app} applies the proposed methods in an empirical setting. Finally, Section \ref{sec:extensions} generalizes the results to conditional-on-observables IV and multiple units per group, and Section \ref{sec:conclusion} concludes. The supplemental appendix contains the technical proofs and additional results and discussions. 


\section{Setup}\label{sec:setup}

Consider a random sample of independent and identically distributed groups indexed by $g=1,\ldots,G$. Spillovers are assumed to occur between units in the same group, but not between units in different groups. I start by considering the case in which each group consists of two identically-distributed units, so that each unit $i$ in group $g$ has one peer. This setup has a a wide range of applications in which groups consist, for example, of roommates in college dormitories \citep{Sacerdote_2001_QJE,Babcock-etal_2015_JEEA,Garlick_2018_AEJ}, spouses \citep{Fletcher-Marksteiner_2017_AEJ,Foos-DeRooij_2017_AJPS}, siblings \citep{Barrera-Osorio-etal_2011_AEJ}, etc. Section \ref{sec:extensions} generalizes this setup to the case of multiple units per group \citep[see][for an example]{Sacarny-etal_2018_JAMA}.

The goal is to study the effect of a treatment on an outcome of interest $Y_{ig}$, allowing for within-group spillovers. The individual treatment status of unit $i$ in group $g$ is denoted by $D_{ig}$, taking values $d\in\{0,1\}$.\footnote{Section \ref{app:multi} in the supplemental appendix generalizes the setup to multiple treatment levels.} For each unit $i$, $D_{jg}$ with $j\ne i$ is the treatment indicator corresponding to unit $i$'s peer. Treatment take-up can be endogenous and is allowed to be arbitrarily correlated with individual characteristics, both observable and unobservable. To address this endogeneity, I assume the researcher has access to a pair of instrumental variables, $(Z_{ig},Z_{jg})$ for unit $i$ and her peer $j$, taking values $(z,z')\in\{0,1\}^2$. These instruments are as-if randomly assigned in a sense formalized below. Borrowing from the literature on imperfect compliance in RCTs, I will often refer to the instruments $(Z_{ig},Z_{jg})$ as ``assignments'', as they indicate whether an individual is assigned (or encouraged) to get the treatment. However, all the results in the paper apply not only to cases in which the researcher has control on the assignment mechanism of $(Z_{ig},Z_{jg})$, as in an encouragement design, but also to cases in which the instruments come from a natural experiment (\citealp[see e.g.][for general discussions]{Angrist-Krueger_2001_JEP,Titiunik_2020_Chapter} and \citealp[][for an application]{Rincke-Traxler_2011_Restat}).

For a given realization of the treatment statuses $(D_{ig},D_{jg})=(d,d')$ and the instruments $(Z_{ig},Z_{jg})=(z,z')$, the potential outcome for unit $i$ in group $g$ is a random variable denoted by $Y_{ig}(d,d',z,z')$. I assume that instruments do not directly affect potential outcomes, an assumption commonly known as the exclusion restriction.

\begin{assumption}[Exclusion restriction]\label{assu:excl_restr}
$Y_{ig}(d,d',z,z')=Y_{ig}(d,d',\tilde{z},\tilde{z}')$ for all $(z,z',\tilde{z},\tilde{z}')$.
\end{assumption}

Under Assumption \ref{assu:excl_restr}, potential outcomes are a function of treatment status only, $Y_{ig}(d,d')$. In this setting, the direct effect of the treatment on unit $i$ given peer's treatment status $d'$ is defined as $Y_{ig}(1,d')-Y_{ig}(0,d')$ and the spillover effect on unit $i$ given own treatment status $d$ is defined as $Y_{ig}(d,1)-Y_{ig}(d,0)$. The observed outcome for unit $i$ in group $g$ is the value of the potential outcome under the observed treatment realization, $Y_{ig}=Y_{ig}(D_{ig},D_{jg})$. 


The possibility of endogenous treatment status introduces an additional channel through which spillovers can materialize. For example, in an encouragement design targeted at couples, unit $i$ may not be offered the incentive to participate in the program, so that $Z_{ig}=0$, but she may hear about the program from her spouse who is assigned the incentive, $Z_{jg}=1$, and decide to participate so $D_{ig}=1$. More generally, treatment status may depend on both own and peer's assignment. To formalize this phenomenon, the potential treatment status for unit $i$ in group $g$ given instruments values $(Z_{ig},Z_{jg})=(z,z')$ will be denoted by $D_{ig}(z,z')$, and the spillover effect on unit $i$'s treatment status is $D_{ig}(z,1)-D_{ig}(z,0)$ for $z=0,1$. The observed treatment status is $D_{ig}(Z_{ig},Z_{jg})$. 

The following assumption formalizes the requirement that instruments are as-if randomly assigned.

\begin{assumption}[Independence]\label{assu:indep}
Let $\mathbf{y}_{ig}=(Y_{ig}(d,d'))_{(d,d')}$ and $\mathbf{\bar{d}}_{ig}=(D_{ig}(z,z'))_{(z,z')}$. Then, $(\mathbf{y}_{ig},\mathbf{\bar{d}}_{ig},\mathbf{y}_{jg},\mathbf{\bar{d}}_{jg}) \indep (Z_{ig},Z_{jg})$.
\end{assumption}

Section \ref{sec:cia} offers an alternative version of this assumption in which independence holds after conditioning on a set of observable covariates.

The different values that $D_{ig}(z,z')$ can take determine each unit's \textit{compliance type}, which indicates how each unit's treatment status responds to different configurations of instruments $(Z_{ig},Z_{jg})$. A careful analysis of compliance types is crucial for defining and identifying causal parameters in this setting: the instruments affect each compliance type in a different way, which in turns determines what parameters can and cannot be identified by harnessing the exogenous variation generated by the instruments, as I discuss next.

\subsection{Compliance Types}\label{sec:distr_comp}

The vector $(D_{ig}(0,0),D_{ig}(0,1),D_{ig}(1,0),D_{ig}(1,1))$, which indicates the unit's treatment status for each possible assignment, determines each unit's compliance type. For example, a unit with $D_{ig}(z,z')=0$ for all $(z,z')$ always refuses the treatment regardless of her own and her peer's assignment. A unit with $D_{ig}(z,z')=1$ for all $(z,z')$ always receives the treatment regardless of her own and her peer's assignment. A unit with $D_{ig}(1,1)=D_{ig}(1,0)=1$ and $D_{ig}(0,1)=D_{ig}(0,0)=0$ only receives the treatment when she is assigned to it, regardless of her peer's assignment, and so on. Without further restrictions, there is a total of 16 different compliance types in the population. To reduce the number of compliance types, I will introduce the following restriction that generalizes the commonly invoked monotonicity assumption to the spillovers case.

\begin{assumption}[Monotonicity]\label{assu:monot}
For all $i$ and $g$, $D_{ig}(1,1)\ge D_{ig}(1,0)\ge D_{ig}(0,1)\ge D_{ig}(0,0)$.
\end{assumption}

This is a strict generalization of the monotonicity assumption in \citet{Angrist-Imbens-Rubin_1996_JASA} and \citet{Kang-Imbens_2016_wp} who require that $D_{ig}(1,1)= D_{ig}(1,0)$ and $D_{ig}(0,1)=D_{ig}(0,0)$. Assumption \ref{assu:monot} reduces the compliance types to five, listed in Table \ref{tab:types} in decreasing order of likelihood of being treated. Always-takers (AT) are units who receive treatment regardless of own and peer treatment assignment. Social-interaction compliers (SC), a term coined by \citet{Duflo-Saez_2003_QJE}, are units who receive the treatment as soon as someone in their group (either themselves or their peer) is assigned to it. Compliers (C) are units that receive the treatment if and only if they are assigned to it. Group compliers (GC) are units who only receive the treatment when their whole group (i.e. both themselves and their peer) is assigned to treatment. Finally, never-takers (NT) are never treated regardless of own and peer's assignment. Note that monotonicity is not testable, as one can only observe one out of the four possible potential treatment statuses, and hence its validity needs to be assessed on a case-by-case basis.

\begin{table}
\caption{Population types}
\begin{center}
\begin{tabular}{ccccc}\label{tab:types}
$D_{ig}(1,1)$ & $D_{ig}(1,0)$ & $D_{ig}(0,1)$ & $D_{ig}(0,0)$ & Type  \\ \hline \hline
1 & 1 & 1 & 1 & Always-taker (AT) \\
1 & 1 & 1 & 0 & Social-interaction complier (SC) \\
1 & 1 & 0 & 0 & Complier (C) \\
1 & 0 & 0 & 0 & Group complier (GC) \\
0 & 0 & 0 & 0 & Never-taker (NT) \\ \hline
\end{tabular}
\end{center}
\end{table}

In what follows, let $\xi_{ig}$ denote a random variable indicating unit $i$'s compliance type, $\xi_{ig}\in\{\text{AT,SC,C,GC,NT}\}$. Also, let $C_{ig}$ denote the event that unit $i$ in group $g$ is a complier, $C_{ig}=\{\xi_{ig}=\text{C}\}$, and similarly for $AT_{ig}=\{\xi_{ig}=\text{AT}\}$, $SC_{ig}=\{\xi_{ig}=\text{SC}\}$ and so on.

\begin{remark}[Monotonicity and ordering]
The ordering in Assumption \ref{assu:monot} is without loss of generality and can be rearranged depending on the context. For example, if the treatment is alcohol consumption and the instrument is a randomly assigned incentive to reduce alcohol intake, the inequalities can be reverted, $D_{ig}(1,1)\le D_{ig}(1,0)\le D_{ig}(0,1)\le D_{ig}(0,0)$. More generally, any ordering is possible provided the same ordering holds for all units in the population.
\end{remark}

\begin{remark}[Connection to multi-valued instruments and treatments]\label{rmk:multi}
The setup in this paper can be mapped into one where the pair $(Z_{ig},Z_{jg})$ is viewed a multi-valued instrument $A_{ig}=2Z_{ig}+Z_{jg}\in\{0,1,2,3\}$ and $(D_{ig},D_{jg})$ is viewed a multi-valued treatment $T_{ig}=2D_{ig}+D_{jg}\in\{0,1,2,3\}$, where $T_{ig}(a)$ denotes potential treatment status under assignment $a$. This setup is analyzed under general conditions by \citet{Heckman-Pinto_2018_ECMA}. As shown in the upcoming sections, the spillovers case introduces specific modeling restrictions and a different way to interpret heterogeneity in treatment effects. In particular, unordered monotonicity \citep[Assumption A-3 in][]{Heckman-Pinto_2018_ECMA} does not hold in this setting. For example, unordered monotonicity requires that either $\1(T_{ig}(1)=3)\ge \1(T_{ig}(2)=3)$ for all $i,g$ or $\1(T_{ig}(1)=3)\le \1(T_{ig}(2)=3)$ for all $i,g$. In my setting, $\1(T_{ig}(1)=3)=D_{ig}(0,1)D_{jg}(1,0)$ and $\1(T_{ig}(2)=3)=D_{ig}(1,0)D_{jg}(0,1)$. Now, in the event $\{AT_{ig},C_{jg}\}$, we have that $D_{ig}(0,1)D_{jg}(1,0)=1>D_{ig}(1,0)D_{jg}(0,1)=0$, whereas in the event $\{C_{ig},AT_{jg}\}$, $D_{ig}(0,1)D_{jg}(1,0)=0<D_{ig}(1,0)D_{jg}(0,1)=1$ and hence the weak inequality does not hold.
\end{remark}

Under Assumptions \ref{assu:excl_restr}, \ref{assu:indep} and \ref{assu:monot}, the marginal distribution of compliance types in the population is identified, as the following proposition shows.

\begin{proposition}[Distribution of compliance types]\label{prop:comp_distr}
Under Assumptions \ref{assu:excl_restr}-\ref{assu:monot},
\begin{align*}
\P[AT_{ig}] &=\E[D_{ig}|Z_{ig}=0,Z_{jg}=0] \\
\P[SC_{ig}]&=\E[D_{ig}|Z_{ig}=0,Z_{jg}=1]-\E[D_{ig}|Z_{ig}=0,Z_{jg}=0]\\
\P[C_{ig}]&=\E[D_{ig}|Z_{ig}=1,Z_{jg}=0]-\E[D_{ig}|Z_{ig}=0,Z_{jg}=1]\\
\P[GC_{ig}]&=\E[D_{ig}|Z_{ig}=1,Z_{jg}=1]-\E[D_{ig}|Z_{ig}=1,Z_{jg}=0]
\end{align*}
and $\P[NT_{ig}]=1-\P[AT_{ig}]-\P[SC_{ig}]-\P[C_{ig}]-\P[GC_{ig}]$. Finally, $\P[AT_{ig},AT_{jg}]=\E[D_{ig}D_{jg}|Z_{ig}=0,Z_{jg}=0]$ and $\P[NT_{ig},NT_{jg}]=\E[(1-D_{ig})(1-D_{jg})|Z_{ig}=1,Z_{jg}=1]$.
\end{proposition}

All the proofs can be found in the supplemental appendix. Proposition \ref{prop:comp_distr} can be used to test for the presence of average spillover effects on treatment status. Note that under Assumption \ref{assu:monot}, $\E[D_{ig}(0,1)-D_{ig}(0,0)]=\P[SC_{ig}]$ and $\E[D_{ig}(1,1)-D_{ig}(1,0)]=\P[GC_{ig}]$, and thus testing for the presence of average spillover effects on treatment status amounts to testing for the presence of social-interaction compliers and group compliers. Because the instruments are as-if randomly assigned, these issues can be analyzed within the framework in \citet{Vazquez-Bare_2017_framework}.

\subsection{Causal Parameters and Estimands of Interest}\label{sec:estimands}

In the presence of spillovers, average direct and spillover effects can be defined as differences between average potential outcomes under different treatment configurations, $\E[Y_{ig}(d,d')-Y_{ig}(\tilde{d},\tilde{d}')]$. When the treatment vector $(D_{ig},D_{jg})$ is randomly assigned, average potential outcomes (and thus treatment effects) are identified by the relationship $\E[Y_{ig}(d,d')]=\E[Y_{ig}|D_{ig}=d,D_{jg}=d']$ \citep[see][and references therein]{Vazquez-Bare_2017_framework}. 

When the treatment is endogenous, average causal effects are generally not point identified. In the absence of spillovers, that is, when $Y_{ig}(d,d')=Y_{ig}(d)$ and $D_{ig}(z,z')=D_{ig}(z)$, the instrument's exogenous variation can be leveraged to identify the local average effect on compliers, $\E[Y_{ig}(1)-Y_{ig}(0)|D_{ig}(1)>D_{ig}(0)]$, emphasizing that average potential outcomes and treatment effects can vary over compliance types, and that the instrument only provides identifying variation on the specific subpopulation whose behavior is affected by the instrument. In the presence of spillovers, without further restrictions, average potential outcomes can depend on both own and peer's compliance types, $\E[Y_{ig}(d,d')|\xi_{ig}=\xi,\xi_{jg}=\xi']$. I refer to these parameters as ``local average potential outcomes''.

The upcoming section shows that, in general, the simultaneous presence of spillovers on treatment status and outcomes can impede identification of causally interpretable parameters even when the instruments are randomly assigned. However, Section \ref{sec:identif} shows that when noncompliance is one-sided, it is possible to identify the average direct effect on compliers, $\E[Y_{ig}(1,0)-Y_{ig}(0,0)|C_{ig}]$ and the average spillover effect on units with compliant peers, $\E[Y_{ig}(0,1)-Y_{ig}(0,0)|C_{jg}]$. Section \ref{sec:extensions} provides an alternative identification assumption that applies to the case of multiple units per group.


\section{Intention-to-Treat Parameters}\label{sec:ITT}

Intention-to-treat (ITT) analysis focuses on the variation in $Y_{ig}$ generated by the instruments. In the absence of spillovers, the ITT estimand $\E[Y_{ig}|Z_{ig}=1]-\E[Y_{ig}|Z_{ig}=0]$ is an attenuated measure of the average treatment effect on compliers, or local average treatment effect (LATE). Furthermore, the LATE can be easily recovered by rescaling the ITT by the proportion of compliers, which is identified under monotonicity and as-if random assignment of the instrument. This section shows that, in the presence of spillovers, the link between ITT parameters and local average effects is much less clear, as the former will conflate multiple potentially different effects into a single number that may be hard to interpret in the presence of treatment effect heterogeneity. 

I will refer to differences in average outcomes changing own instrument leaving the peer's instrument fixed as \textit{direct} ITT parameters, $\E[Y_{ig}|Z_{ig}=1,Z_{jg}=z']-\E[Y_{ig}|Z_{ig}=0,Z_{jg}=z']$, and differences fixing own instrument and varying the peer's instrument as \textit{indirect} or \textit{spillover} ITT parameters, $\E[Y_{ig}|Z_{ig}=z,Z_{jg}=1]-\E[Y_{ig}|Z_{ig}=z,Z_{jg}=0]$. Finally, the \textit{total} ITT is defined as $\E[Y_{ig}|Z_{ig}=1,Z_{jg}=1]-\E[Y_{ig}|Z_{ig}=0,Z_{jg}=0]$.

The following result links the direct ITT estimand to potential outcomes. In what follows, the notation $\{C_{ig},SC_{ig}\}\times\{AT_{jg}\}$ refers to the event $(C_{ig}\cap AT_{jg})\cup (SC_{ig}\cap AT_{jg})$, that is, unit $j$ is an always-taker and unit $i$ can be a complier or a social complier. Similarly, $\{C_{ig},SC_{ig}\}\times \{C_{jg},GC_{jg},NT_{jg}\}$ represents all the combinations in which unit $i$ is a complier or a social complier and unit $j$ is a complier, a group complier or a never-taker, and so on.

\begin{lemma}[Direct ITT effects]\label{lemma:direct_ITT}
Under Assumptions \ref{assu:excl_restr}-\ref{assu:monot},
\begin{align*}
\E[Y_{ig}&|Z_{ig}=1,Z_{jg}=0]-\E[Y_{ig}|Z_{ig}=0,Z_{jg}=0]=\\
&\E[Y_{ig}(1,0)-Y_{ig}(0,0)|\{C_{ig},SC_{ig}\}\times \{C_{jg},GC_{jg},NT_{jg}\}]\P[\{C_{ig},SC_{ig}\}\times \{C_{jg},GC_{jg},NT_{jg}\}]\\
+&\E[Y_{ig}(1,1)-Y_{ig}(0,0)|\{C_{ig},SC_{ig}\}\times\{SC_{jg}\}]\P[\{C_{ig},SC_{ig}\}\times\{SC_{jg}\}]\\
+&\E[Y_{ig}(1,1)-Y_{ig}(0,1)|\{C_{ig},SC_{ig}\}\times\{AT_{jg}\}]\P[\{C_{ig},SC_{ig}\}\times\{AT_{jg}\}]\\
+&\E[Y_{ig}(0,1)-Y_{ig}(0,0)|\{GC_{ig},NT_{ig}\}\times\{SC_{jg}\}]\P[\{GC_{ig},NT_{ig}\}\times\{SC_{jg}\}]\\
+&\E[Y_{ig}(1,1)-Y_{ig}(1,0)|AT_{ig},SC_{jg}]\P[AT_{ig},SC_{jg}].
\end{align*}
\end{lemma}

The corresponding results for the indirect ITT and the total ITT are analogous, and are presented in Section \ref{app:addresults} of the supplemental appendix to conserve space. 

To interpret the above result, consider the effect of switching $Z_{ig}$ from 0 to 1, leaving $Z_{jg}$ fixed at zero. First, if unit $i$ is either a complier or a social complier, switching $Z_{ig}$ from 0 to 1 will change her treatment status $D_{ig} $ from 0 to 1. This case corresponds to the first three expectations on the right-hand side of Lemma \ref{lemma:direct_ITT}. Now, if unit $j$ is a complier, a group complier or a never-taker, her observed treatment status would be $D_{jg}=0$. Hence, in these cases, switching $Z_{ig}$ from 0 to 1 while leaving $Z_{jg}$ fixed at zero would let us observe $Y_{ig}(1,0)-Y_{ig}(0,0)$. This corresponds to the first expectation on the right-hand side of Lemma \ref{lemma:direct_ITT}. On the other hand, if unit $j$ was a social complier, switching $Z_{ig}$ from 0 to 1 would push her to get the treatment, and hence in this case we would see $Y_{ig}(1,1)-Y_{ig}(0,0)$. This case corresponds to the second expectation on the right-hand side of Lemma \ref{lemma:direct_ITT}. If instead unit $j$ was an always-taker, she would be treated in both scenarios, so we would see $Y_{ig}(1,1)-Y_{ig}(0,1)$ (third expectation of the above display). Next, suppose unit $i$ was a group complier or a never-taker. Then, switching $Z_{ig}$ from 0 to 1 would not affect her treatment status, which would be fixed at 0, but it would affect unit $j$'s treatment status if she is a social complier. This case is shown in the fourth expectation on the right-hand side of Lemma \ref{lemma:direct_ITT}. Finally, if unit $i$ was an always-taker, her treatment status would be fixed at 1 but her peer's treatment status would switch from 0 to 1 if unit $j$ was a social complier. This case is shown in the last expectation on the right-hand side of Lemma \ref{lemma:direct_ITT}.

Hence the direct ITT effect is averaging five different treatment effects, $Y_{ig}(1,0)-Y_{ig}(0,0)$, $Y_{ig}(1,1)-Y_{ig}(0,0)$, $Y_{ig}(1,1)-Y_{ig}(1,0)$, $Y_{ig}(0,1)-Y_{ig}(0,0)$, and $Y_{ig}(1,1)-Y_{ig}(0,1)$, each one over different combinations of compliance types. Therefore, Lemma \ref{lemma:direct_ITT} shows that, even when fixing the peer's assignment, the ITT parameter is unable to isolate direct and indirect effects, which blurs its link to causal effects.

In some contexts, ITT parameters are deemed policy relevant as they measure the ``effect'' of offering the treatment or making the treatment available, as opposed to the effect of the treatment itself \citep{Abadie-Cattaneo_2018_Annurev}. This interpretation is based on the fact that, without spillovers, the ITT is $\E[Y_{ig}|Z_{ig}=1]-\E[Y_{ig}|Z_{ig}=0]=\E[Y_{ig}(1)-Y_{ig}(0)|C_{ig}]\P[C_{ig}]$ which is the local average treatment effect, down-weighted by the compliance rate. In particular, this well-known fact has three implications that facilitate the interpretation of the ITT as a policy-relevant parameter: (i) it has the same sign as the LATE, and it equals zero if and only if the LATE is zero (unless the instrument is completely irrelevant) (ii) it is a lower bound for the LATE (in absolute value) and (iii) it is proportional to the LATE, so it can be easily rescaled to recover the LATE.

Lemma \ref{lemma:direct_ITT} shows that the close link between the ITT and the LATE breaks down in the presence of spillovers. First, the ITT now combines multiple average direct and spillover effects which can have different signs and magnitudes. In particular, the ITT could be zero even if all treatment effects are non-zero. Second, for this same reason, the ITT is no longer a lower bound for any of the direct or spillover effects. Finally, rescaling the ITT by the first stage $\E[D_{ig}|Z_{ig}=1,Z_{jg}=0]-\E[D_{ig}|Z_{ig}=0,Z_{jg}=0]$ does not recover a treatment effect.\footnote{Notice that rescaling the ITT by the first stage recovers the estimand from a 2SLS regression instrumenting $D_{ig}$ with $Z_{ig}$ conditional on $Z_{jg}=0$.} Specifically, the weights from the direct ITT sum to $\P[C_{ig}]+\P[SC_{ig}]+\P[SC_{ig},GC_{jg}]+\P[SC_{ig},NT_{jg}]+\P[SC_{ig},AT_{jg}]$, whereas $\E[D_{ig}|Z_{ig}=1,Z_{jg}=0]-\E[D_{ig}|Z_{ig}=0,Z_{jg}=0]=\P[C_{ig}]+\P[SC_{ig}]$ from Proposition \ref{prop:comp_distr}. 

\begin{remark}[Spillovers and instrument validity]
Another way to interpret the result in Lemma \ref{lemma:direct_ITT} is to think of spillovers in treatment take-up as violating instrument validity. Since $D_{jg}$ is a function of $Z_{ig}$, the instrument $Z_{ig}$ can affect the outcome $Y_{ig}$ not only through the variable it is instrumenting, $D_{ig}$, but also through another variable $D_{jg}$. Thus, spillovers on treatment take-up may render an instrument invalid even when the instrument would have been valid in the absence of spillovers. This fact shows that identification of causal parameters based on $(Z_{ig},Z_{jg})$ will require further assumptions, as discussed in the next section.
\end{remark}

In all, this section shows that ITT parameters are generally not a useful measure of average direct and spillover effects. Instead of focusing on ITT parameters, which rely exclusively on variation generated by the instruments $(Z_{ig},Z_{jg})$, an alternative approach for identification is to exploit the combined variation in $(Z_{ig},Z_{jg},D_{ig},D_{jg})$. \citet{Imbens-Rubin_1997_ReStud} show that, in the absence of spillovers, this approach allows for separate point identification of average (or distributions of) potential outcomes for compliers. This approach, however, breaks down in the presence of spillovers. The reason is that, without further assumptions, the possible combinations of treatment and instrument values are not enough to disentangle all the different compliance types. Further details on this issue are provided in Section \ref{app:identif_monot} of the supplemental appendix. In what follows, I show that point identification can be restored when the degree of noncompliance is restricted.


\section{Identification Under One-sided Noncompliance}\label{sec:identif}

This section shows that identification of some causal parameters can be achieved by limiting the amount of noncompliance. I will analyze the case in which noncompliance is one-sided. One-sided noncompliance (OSN) refers to the case in which individual deviations from their assigned treatment, $D_{ig}\ne Z_{ig}$, can only occur in one direction. 

In many applications, units who are not assigned to treatment are unable to get the treatment through other channels. For example, consider a voter turnout experiment in which individuals in two-voter households are randomly assigned to receive a telephone call encouraging them to vote \citep[as in][]{Foos-DeRooij_2017_AJPS}. In this case, units that are assigned $Z_{ig}=1$ may fail to receive the actual phone call (for example, because they refuse to pick up the phone), in which case $Z_{ig}=1$ and $D_{ig}=0$, but whenever a unit is assigned $Z_{ig}=0$, this automatically implies $D_{ig}=0$. More generally, one-sided noncompliance is common when the experimenter is the only provider of a treatment \citep{Abadie-Cattaneo_2018_Annurev}. I formalize this case as follows.

\begin{assumption}[One-sided Noncompliance]\label{assu:onesided}
$\P[D_{ig}=1|Z_{ig}=0]=0$.
\end{assumption}

Under Assumption \ref{assu:monot} (monotonicity), one-sided noncompliance implies the absence of always-takers and social-interaction compliers. Notice that Assumption \ref{assu:onesided} is testable. In what follows, all the results focus on identifying the expectation of potential outcomes, but these results immediately generalize to identification of marginal distributions of potential outcomes by replacing $Y_{ig}$ by $\1(Y_{ig}\le y)$.

\begin{proposition}[Local average potential outcomes under OSN]\label{prop:lapo_osnc}
Under Assumptions \ref{assu:excl_restr}-\ref{assu:onesided}, the following equalities hold:
\begin{align*}
\E[Y_{ig}(0,0)]&=\E[Y_{ig}|Z_{ig}=0,Z_{jg}=0]\\
\E[Y_{ig}(1,0)|C_{ig}]\P[C_{ig}]&=\E[Y_{ig}D_{ig}|Z_{ig}=1,Z_{jg}=0]\\
\E[Y_{ig}(0,1)|C_{jg}]\P[C_{jg}]&=\E[Y_{ig}D_{jg}|Z_{ig}=0,Z_{jg}=1]\\
\E[Y_{ig}(0,0)|C_{ig}]\P[C_{ig}]&=\E[Y_{ig}|Z_{ig}=0,Z_{jg}=0]-\E[Y_{ig}(1-D_{ig})|Z_{ig}=1,Z_{jg}=0]\\
\E[Y_{ig}(0,0)|C_{jg}]\P[C_{jg}]&=\E[Y_{ig}|Z_{ig}=0,Z_{jg}=0]-\E[Y_{ig}(1-D_{jg})|Z_{ig}=0,Z_{jg}=1]\\
\E[Y_{ig}(0,0)|NT_{ig},NT_{jg}]\P[NT_{ig},NT_{jg}]&=\E[Y_{ig}(1-D_{ig})(1-D_{jg})|Z_{ig}=1,Z_{jg}=1]
\end{align*}
where $\P[NT_{ig},NT_{jg}]=\E[(1-D_{ig})(1-D_{jg})|Z_{ig}=1,Z_{jg}=1]$.
\end{proposition}

Combined with Proposition \ref{prop:comp_distr}, the above result shows which local average potential outcomes can be identified by exploiting variation in the observed treatment status and instruments $(D_{ig},D_{jg},Z_{ig},Z_{jg})$. Proposition \ref{prop:lapo_osnc} has the following implication.

\begin{corollary}[Local average direct and spillover effects under OSN]\label{coro:late_osnc}
Under Assumptions \ref{assu:excl_restr}-\ref{assu:onesided}, if $\P[C_{ig}]>0$,
\begin{align*}
\E[Y_{ig}(1,0)-Y_{ig}(0,0)|C_{ig}]&=\frac{\E[Y_{ig}|Z_{ig}=1,Z_{jg}=0]-\E[Y_{ig}|Z_{ig}=0,Z_{jg}=0]}{\E[D_{ig}|Z_{ig}=1,Z_{jg}=0]}
\end{align*}
and
\begin{align*}
\E[Y_{ig}(0,1)-Y_{ig}(0,0)|C_{jg}]&=\frac{\E[Y_{ig}|Z_{ig}=0,Z_{jg}=1]-\E[Y_{ig}|Z_{ig}=0,Z_{jg}=0]}{\E[D_{jg}|Z_{ig}=0,Z_{jg}=1]}.
\end{align*}
\end{corollary}

In the above result, $\E[Y_{ig}(1,0)-Y_{ig}(0,0)|C_{ig}]$ represents the average direct effect on compliers with untreated peers and $\E[Y_{ig}(0,1)-Y_{ig}(0,0)|C_{jg}]$ is the average effect on untreated units with compliant peers. See Section \ref{sec:emp_app} for a detailed discussion on these estimands in the context of an empirical application. Section \ref{app:multi} generalizes these results to the case of multiple treatment levels.

In addition to identifying these treatment effects, Proposition \ref{prop:lapo_osnc} can be used to assess whether average baseline potential outcomes vary across own and peer compliance types, as the following corollary shows. In what follows, $C^c_{ig}$ represents the event in which unit $i$ is a non-complier, that is, $C^c_{ig}=NT_{ig}\cup GC_{ig}$.

\begin{corollary}[Heterogeneity over compliance types]\label{coro:het_osnc}
Under Assumptions \ref{assu:excl_restr}-\ref{assu:onesided}, if $0<\P[C_{ig}]<1$,
\begin{multline*}
\E[Y_{ig}(0,0)|C_{ig}]-\E[Y_{ig}(0,0)|C_{ig}^c]=\\
\left\{\E[Y_{ig}|Z_{ig}=0,Z_{jg}=0]-\frac{\E[Y_{ig}(1-D_{ig})|Z_{ig}=1,Z_{jg}=0]}{1-\E[D_{ig}|Z_{ig}=1,Z_{jg}=0]}\right\}\frac{1}{\E[D_{ig}|Z_{ig}=1,Z_{ig}=0]}
\end{multline*}
and
\begin{multline*}
\E[Y_{ig}(0,0)|C_{jg}]-\E[Y_{ig}(0,0)|C_{jg}^c]=\\
\left\{\E[Y_{ig}|Z_{ig}=0,Z_{jg}=0]-\frac{\E[Y_{ig}(1-D_{jg})|Z_{ig}=0,Z_{jg}=1]}{1-\E[D_{jg}|Z_{ig}=0,Z_{jg}=1]}\right\}\frac{1}{\E[D_{jg}|Z_{ig}=0,Z_{ig}=1]}.
\end{multline*}
\end{corollary}

The first term in the above corollary, $\E[Y_{ig}(0,0)|C_{ig}]-\E[Y_{ig}(0,0)|C_{ig}^c]$, is the difference in the average baseline outcome $Y_{ig}(0,0)$ between compliers and non-compliers, whereas $\E[Y_{ig}(0,0)|C_{jg}]-\E[Y_{ig}(0,0)|C_{jg}^c]$ is the difference in average baseline potential outcomes among units with compliant and non-compliant peers. These differences can be used to determine whether average baseline potential outcomes vary with own and peers' compliance types.

\subsection{Testing Instrument Validity}\label{sec:spectest}

While identification assumptions are generally not testable, the existing literature on instrumental variables has provided ways to indirectly assess their validity by deriving testable implications \citep{Balke-Pearl_1997_JASA,Imbens-Rubin_1997_ReStud,Kitagawa_2015_ECMA}. Based on these results, the following proposition provides a way to test for instrument validity in the presence of spillovers.

\begin{proposition}[Instrument Validity]\label{prop:spec}
Under Assumptions \ref{assu:excl_restr}-\ref{assu:onesided}, for any Borel set $\mathcal{Y}$,
\[\P[Y_{ig}\in\mathcal{Y},D_{ig}=0|Z_{ig}=1,Z_{jg}=0]-\P[Y_{ig}\in\mathcal{Y}|Z_{ig}=0,Z_{jg}=0]\le 0\]
and 
\[\P[Y_{ig}\in\mathcal{Y},D_{jg}=0|Z_{ig}=0,Z_{jg}=1]-\P[Y_{ig}\in\mathcal{Y}|Z_{ig}=0,Z_{jg}=0]\le 0.\]
\end{proposition}

Because all the terms in Proposition \ref{prop:spec} only involve observable variables, estimating them can provide a way to assess the validity of the identification assumptions. Estimation and inference for these objects can be conducted using the local regression distribution methods in \citet{Cattaneo-Jansson-Ma_2020_JoE} and \citet{Cattaneo-Jansson-Ma_2020_JSS}. 


\section{Estimation and Inference}\label{sec:estimation}

This section discusses estimation and inference for the causal effects in Corollary \ref{coro:late_osnc}. In what follows, let $\mathbf{y}_g=(Y_{1g},Y_{2g})'$, $\mathbf{Z}_g=(Z_{1g},Z_{2g})'$ and $\mathbf{D}_g=(D_{1g},D_{2g})'$. 

\begin{assumption}[Sampling and moments]\label{assu:sampling} $ $
\begin{enumerate}[(a)]
\item $(\mathbf{y}_g',\mathbf{Z}_g',\mathbf{D}_g')_{g=1}^G$ are independent and identically distributed across $g$.
\item For each $g$, $(Y_{1g},Z_{1g},D_{1g})$ and $(Y_{2g},Z_{2g},D_{2g})$ are identically distributed, not necessarily independent.
\item $\E[Y_{ig}^4]< \infty$.
\end{enumerate} 
\end{assumption}

Part (a) of Assumption \ref{assu:sampling} states that the researcher has access to a random sample of iid groups. Part (b) states that observations within each group are identically distributed, but allows for an unrestricted correlation structure within groups. Part (c) is a standard regularity condition to ensure the appropriate moments are bounded.

The average direct and spillover effects in Corollary \ref{coro:late_osnc} can be estimated straightforwardly by replacing the right-hand side of each expression by their sample analog. Because these estimators are IV estimators using a binary instrument and a binary treatment on a specific subsample, I will refer to these estimators as \textit{conditional Wald estimators}. More precisely, I define conditional Wald estimators and their corresponding cluster-robust variance estimator as follows.

\begin{definition}[Direct conditional Wald estimator]\label{def:cwr1}
Let $\mathbf{\tilde{z}}_{ig}=(1,Z_{ig})'$, $\mathbf{\tilde{d}}_{ig}=(1,D_{ig})'$, $\mathbf{\tilde{z}}_g=(\mathbf{\tilde{z}}_{1g}',\mathbf{\tilde{z}}_{2g}')'$. The direct conditional Wald estimator $\boldsymbol{\hat{\delta}}=(\hat{\delta}_0,\hat{\delta}_1)'$ is defined as the estimator from the 2SLS regression of $Y_{ig}$ on an intercept and $D_{ig}$ using $Z_{ig}$ as an instrument, on the subsample of observations with $Z_{jg}=0$, that is,
\[\boldsymbol{\hat{\delta}}=\begin{bmatrix}
\hat{\delta}_0 \\
\hat{\delta}_1
\end{bmatrix}= \left(\sum_g \mathbf{\tilde{z}}_g'\mathbf{\tilde{w}}_g\mathbf{\tilde{d}}_g\right)^{-1}\sum_g \mathbf{\tilde{z}}_g'\mathbf{\tilde{w}}_g\mathbf{y}_g\]
whenever $\sum_{g}\sum_{i}(1-Z_{ig})(1-Z_{jg})>0$, $\sum_{g}\sum_{i}Z_{ig}(1-Z_{jg})>0$ and $\sum_{g}\sum_{i}D_{ig}Z_{ig}(1-Z_{jg})>0$, where
\[\mathbf{\tilde{w}}_g=\begin{bmatrix}
1-Z_{2g} & 0 \\
0 & 1-Z_{1g}
\end{bmatrix}.\]
The cluster-robust variance estimator for $\boldsymbol{\hat{\delta}}$ is:
\[\mathbf{\hat{V}}_{\mathsf{cr}}(\boldsymbol{\hat{\delta}})=\left(\sum_g \mathbf{\tilde{z}}_g'\mathbf{\tilde{w}}_g\mathbf{\tilde{d}}_g\right)^{-1}\sum_g \mathbf{\tilde{z}}_g'\mathbf{\tilde{w}}_g \mathbf{\hat{u}}_g\mathbf{\hat{u}}_g'\mathbf{\tilde{w}}_g\mathbf{\tilde{z}}_g\left(\sum_g \mathbf{\tilde{d}}_g'\mathbf{\tilde{w}}_g\mathbf{\tilde{z}}_g\right)^{-1}\]
where $\hat{u}_{ig}=Y_{ig}-\mathbf{\tilde{d}}_{ig}'\boldsymbol{\hat{\delta}}$ and $\mathbf{\hat{u}}_g=(\hat{u}_{1g},\hat{u}_{2g})'$.
\end{definition}

\begin{definition}[Indirect conditional Wald estimator]\label{def:cwr2}
Let $\mathbf{\check{z}}_{ig}=(1,Z_{jg})'$, $\mathbf{\check{d}}_{ig}=(1,D_{jg})'$, $\mathbf{\check{z}}_g=(\mathbf{\check{z}}_{1g}',\mathbf{\check{z}}_{2g}')'$. The indirect conditional Wald estimator $\boldsymbol{\hat{\lambda}}=(\hat{\lambda}_0,\hat{\lambda}_1)'$ is defined as the estimator from the 2SLS regression of $Y_{ig}$ on an intercept and $D_{jg}$ using $Z_{jg}$ as an instrument, on the subsample of observations with $Z_{ig}=0$, that is,
\[\boldsymbol{\hat{\lambda}}=\begin{bmatrix}
\hat{\lambda}_0 \\
\hat{\lambda}_1
\end{bmatrix}= \left(\sum_g \mathbf{\check{z}}_g'\mathbf{\check{w}}_g\mathbf{\check{d}}_g\right)^{-1}\sum_g \mathbf{\check{z}}_g'\mathbf{\check{w}}_g\mathbf{y}_g\]
whenever $\sum_{g}\sum_{i}(1-Z_{ig})(1-Z_{jg})>0$, $\sum_{g}\sum_{i}Z_{jg}(1-Z_{ig})>0$ and $\sum_{g}\sum_{i}D_{jg}Z_{jg}(1-Z_{ig})>0$, where
\[\mathbf{\check{w}}_g=\begin{bmatrix}
1-Z_{1g} & 0 \\
0 & 1-Z_{2g}
\end{bmatrix}.\]
The cluster-robust variance estimator for $\boldsymbol{\hat{\lambda}}$ is:
\[\mathbf{\hat{V}}_{\mathsf{cr}}(\boldsymbol{\hat{\lambda}})=\left(\sum_g \mathbf{\check{z}}_g'\mathbf{\check{w}}_g\mathbf{\check{d}}_g\right)^{-1}\sum_g \mathbf{\check{z}}_g'\mathbf{\check{w}}_g \boldsymbol{\hat{\nu}}_g\boldsymbol{\hat{\nu}}_g'\mathbf{\check{w}}_g\mathbf{\check{z}}_g\left(\sum_g \mathbf{\check{d}}_g'\mathbf{\check{w}}_g\mathbf{\check{z}}_g\right)^{-1}\]
where $\hat{\nu}_{ig}=Y_{ig}-\mathbf{\check{d}}_{ig}'\boldsymbol{\hat{\lambda}}$ and $\boldsymbol{\hat{\nu}}_g=(\hat{\nu}_{1g},\hat{\nu}_{2g})'$.
\end{definition}

An alternative estimation strategy in a setting with multiple instruments and multiple endogenous variables is to combine all regressors and instruments into a single 2SLS regression. In a constant coefficients setting, this estimation strategy yields consistent and asymptotically normal estimators. However, these features do not generally extend to a setting with heterogeneous effects. In what follows I show that, under one-sided noncompliance, if the 2SLS regression combining both instruments and endogenous variables is fully saturated, the resulting estimators and cluster-robust standard errors are in fact equivalent to the conditional Wald estimators. I start by defining the fully-saturated 2SLS regression estimator as follows.

\begin{definition}[Saturated 2SLS regression]\label{def:2SLS}
Let $\mathbf{z}_{ig}=(1,Z_{ig},Z_{jg},Z_{ig}Z_{jg})'$, $\mathbf{d}_{ig}=(1,D_{ig},D_{jg},D_{ig}D_{jg})'$, $\mathbf{z}_g=(\mathbf{z}_{1g}',\mathbf{z}_{2g}')'$. The saturated 2SLS regression estimator $\boldsymbol{\hat{\beta}}=(\hat{\beta}_0,\hat{\beta}_1,\hat{\beta}_2,\hat{\beta}_3)'$ is defined as the estimator from the 2SLS regression of $Y_{ig}$ on an intercept, $D_{ig}$, $D_{jg}$ and $D_{ig}D_{jg}$ using $Z_{ig}$, $Z_{jg}$ and $Z_{ig}Z_{jg}$ as instruments on the full sample, that is,
\[\boldsymbol{\hat{\beta}}=\begin{bmatrix}
\hat{\beta}_0 \\
\hat{\beta}_1 \\
\hat{\beta}_2 \\
\hat{\beta}_3
\end{bmatrix}= \left(\sum_g \mathbf{z}_g'\mathbf{d}_g\right)^{-1}\sum_g \mathbf{z}_g'\mathbf{y}_g.\]
whenever $\sum_{g}\sum_{i}(1-Z_{ig})(1-Z_{jg})>0$, $\sum_{g}\sum_{i}Z_{ig}(1-Z_{jg})>0$, $\sum_{g}\sum_{i}Z_{ig}Z_{jg}>0$, $\sum_{g}\sum_{i}D_{ig}Z_{ig}(1-Z_{jg})>0$ and $\sum_{g}\sum_{i}D_{ig}D_{jg}Z_{ig}Z_{jg}>0$. The cluster-robust variance estimator for $\boldsymbol{\hat{\beta}}$ is:
\[\mathbf{\hat{V}}_{\mathsf{cr}}(\boldsymbol{\hat{\beta}})=\left(\sum_g \mathbf{z}_g'\mathbf{d}_g\right)^{-1}\sum_g \mathbf{z}_g'\boldsymbol{\hat{\varepsilon}}_g\boldsymbol{\hat{\varepsilon}}_g'\mathbf{z}_g\left(\sum_g \mathbf{d}_g'\mathbf{z}_g\right)^{-1}\]
where $\hat{\varepsilon}_{ig}=Y_{ig}-\mathbf{d}_{ig}'\boldsymbol{\hat{\beta}}$ and $\boldsymbol{\hat{\varepsilon}}_g=(\hat{\varepsilon}_{1g},\hat{\varepsilon}_{2g})'$.
\end{definition}

The following theorem establishes the equivalence between conditional Wald estimators and the saturated 2SLS estimators.

\begin{theorem}[Equivalence between conditional Wald and 2SLS]\label{thm:2SLS}
Consider the estimators $\boldsymbol{\hat{\delta}}$, $\boldsymbol{\hat{\lambda}}$ and $\boldsymbol{\hat{\beta}}$ from Definitions \ref{def:cwr1}, \ref{def:cwr2} and $\ref{def:2SLS}$. Suppose that Assumptions \ref{assu:excl_restr}-\ref{assu:sampling} hold, and that $\sum_{g}\sum_{i}(1-Z_{ig})(1-Z_{jg})>0$, $\sum_{g}\sum_{i}Z_{ig}(1-Z_{jg})>0$, $\sum_{g}\sum_{i}Z_{ig}Z_{jg}>0$, $\sum_{g}\sum_{i}D_{ig}Z_{ig}(1-Z_{jg})>0$ and $\sum_{g}\sum_{i}D_{ig}D_{jg}Z_{ig}Z_{jg}>0$. Then, $\hat{\delta}_0=\hat{\lambda}_0=\hat{\beta}_0$, $\hat{\delta}_1=\hat{\beta}_1$, $\hat{\lambda}_1=\hat{\beta}_2$
and $\mathbf{\hat{V}}_{\mathsf{cr},11}(\boldsymbol{\hat{\delta}})=\mathbf{\hat{V}}_{\mathsf{cr},11}(\boldsymbol{\hat{\lambda}})=\mathbf{\hat{V}}_{\mathsf{cr},11}(\boldsymbol{\hat{\beta}})$, $\mathbf{\hat{V}}_{\mathsf{cr},22}(\boldsymbol{\hat{\delta}})=\mathbf{\hat{V}}_{\mathsf{cr},22}(\boldsymbol{\hat{\beta}})$ and $\mathbf{\hat{V}}_{\mathsf{cr},22}(\boldsymbol{\hat{\lambda}})=\mathbf{\hat{V}}_{\mathsf{cr},33}(\boldsymbol{\hat{\beta}})$.
\end{theorem}

In what follows let ``$\to_\P$'' denote convergence in probability and ``$\to_\mathcal{D}$'' denote convergence in distribution. The following result shows that the 2SLS are consistent and asymptotically normal. This result is standard in 2SLS and included only for completion.

\begin{lemma}[Consistency and asymptotic normality]\label{lemma:asynorm}
Under Assumptions \ref{assu:excl_restr}-\ref{assu:sampling}, if $\P[Z_{ig}=0,Z_{jg}=0]>0$, $\P[Z_{ig}=1,Z_{jg}=0]>0$, $\P[Z_{ig}=1,Z_{jg}=1]>0$, $\E[D_{ig}|Z_{ig}=1,Z_{jg}=0]>0$ and $\E[D_{ig}D_{jg}|Z_{ig}=1,Z_{jg}=1]>0$, then as $G\to\infty$,
\[\boldsymbol{\hat{\beta}}\to_\P\boldsymbol{\beta}=\begin{bmatrix}
\E[Y_{ig}(0,0)]\\
\E[Y_{ig}(1,0)-Y_{ig}(0,0)|C_{ig}]\\
\E[Y_{ig}(0,1)-Y_{ig}(0,0)|C_{jg}]\\
\beta_3
\end{bmatrix}\]
and
\[\sqrt{2G}(\boldsymbol{\hat{\beta}}-\boldsymbol{\beta})\to_\mathcal{D}\mathcal{N}(\mathbf{0},\mathbf{V}),\quad \mathbf{V}=\E[\mathbf{z}_g'\mathbf{d}_g]^{-1}\E[ \mathbf{z}_g'\boldsymbol{\varepsilon}_g\boldsymbol{\varepsilon}_g'\mathbf{z}_g]\E[\mathbf{d}_g'\mathbf{z}_g]^{-1}\]
where $\varepsilon_{ig}=Y_{ig}-\mathbf{d}_g'\boldsymbol{\beta}$ and $\boldsymbol{\varepsilon}_g=(\varepsilon_{1g},\varepsilon_{2g})'$. In addition, $2G\mathbf{\hat{V}}_{\mathsf{cr}}(\boldsymbol{\hat{\beta}})\to_\P \mathbf{V}$.
\end{lemma}
The interaction population coefficient $\beta_3$ does not have a direct causal interpretation, and its exact formula is given in the proof in the supplemental appendix.

\begin{remark}[Interaction terms]
Although the coefficient on $\hat{\beta}_3$ corresponding to the interaction term $D_{ig}D_{jg}$ does not have a direct causal interpretation, the terms $D_{ig}D_{jg}$ and $Z_{ig}Z_{jg}$ need to be included to ensure the equivalence of the estimators by saturating the model. If $\sum_{g}\sum_{i}Z_{ig}Z_{jg}=0$, under imperfect compliance $D_{ig}D_{jg}=0$ and the results from Theorem \ref{thm:2SLS} hold after excluding the interaction terms $D_{ig}D_{jg}$ and $Z_{ig}Z_{jg}$ from the estimation procedure.
\end{remark}

Notice that the magnitudes of interest defined in Corollary \ref{coro:het_osnc} and Proposition \ref{prop:spec} cannot be written as 2SLS estimators, and hence Theorem \ref{thm:2SLS} and Lemma \ref{lemma:asynorm} do not apply directly. However, all these parameters are nonlinear functions of sample means, and hence consistency and asymptotic normality follows under standard conditions. I provide further details on estimation and inference for these parameters in the supplemental appendix.

\subsection{Weak-Instrument-Robust Inference}\label{sec:weakiv}

When non-compliance is severe, the proportion of compliers can be close to zero and, as a result, instruments may be weak. In such cases, the estimators analyzed above may have poor finite-sample properties, and inference based on the normal approximation may be unreliable. The 2SLS literature has provided several alternatives to conduct inference that is robust to weak instruments \citep[see][for a recent review]{Andrews-stock-Sun_2019_Annurev}. In particular, Fieller or Anderson-Rubin (AR) confidence intervals have been shown to provide correct coverage regardless of the strength of the instruments. 

By Theorem \ref{thm:2SLS}, the average direct and spillover effects can be estimated by two separate regressions that involve a single binary instrument and a single binary endogenous variable (the conditional Wald estimators), which give the same estimates and standard errors as the saturated 2SLS regression. The advantage of the conditional Wald estimators is that their corresponding AR confidence intervals can be obtained by solving a quadratic inequality as shown by \citet{Dufour-Taamouti_2005_ECMA} and \citet{Mikusheva_2010_JoE}, without the need of computationally intensive grid searches or projection methods. Section \ref{app:AR} provides further details on how to construct AR confidence intervals in this context.


\section{Application: Spillovers in Voting Behavior}\label{sec:emp_app}

In this section I illustrate the above results using data from a randomized experiment on voter mobilization conducted by \citet{Foos-DeRooij_2017_AJPS}. The goal of their study is to assess if political discussions within close social networks such as the household have an effect on voter turnout, and, if so, in what direction.

To this end, the authors conducted a randomized experiment in which two-voter households in Birmingham, UK were randomly assigned to receive a telephone message providing information and encouraging people to vote on the West Midlands Police and Crime Commissioner (PCC) 2012 election. A sample of 5,190 two-voter households with landline numbers were divided into treatment and control households, and within the households assigned to the treatment, only one household member was randomly selected to receive the telephone message.\footnote{The experiment had two different treatment intensities, which I pool here for illustration purposes. Section \ref{app:multi_emp} in the supplemental appendix provides a detailed analysis of the multi-level treatment.}

Because the telephone message is delivered by landline, this type of experiments is usually subject to severe rates of nonresponse, since individuals assigned to treatment are likely to be unavailable, refuse to participate, may have moved or their phone numbers can be outdated or wrong. For these and other reasons, it is common to find compliance rates below 50 percent \citep{Gerber-Green_2000_AJPS}. In the experiment described here, the response rate among individuals assigned to receive the message is about 45 percent. To account for the potential endogeneity of this type of noncompliance, the randomized treatment assignment can be used as an instrument for actual treatment receipt. More precisely, for each household $g$, let $(Z_{ig},Z_{jg})$ be the randomized treatment assignment for each unit, where $Z_{ig}=1$ if individual $i$ is randomly assigned to receive the phone call. Let $(D_{ig},D_{jg})$ be the treatment indicators, where $D_{ig}=1$ if individual $i$ actually receives the phone message. Finally, the outcome of interest $Y_{ig}$, voter turnout, equals 1 if individual $i$ voted in the election.

In this experiment, noncompliance is one-sided, as units assigned to treatment can fail to receive the phone call, but units assigned to control do not receive it. Since only one member of each treated household was selected to receive the call, we also have that $\P[Z_{ig}=1,Z_{jg}=1]=0$. Given this experimental design, the first stage reduces to estimating $\E[D_{ig}|Z_{ig}=1,Z_{jg}=0]=\E[D_{ig}|Z_{ig}=1]$. The estimated coefficient is $0.451$, significantly different from zero at the one percent level and with an $F$-statistic of $1759.03$, which suggests a strong instrument.

The estimation results are shown in Table \ref{tab:results}. Column (1) shows the naive estimates obtained by ignoring the presence of spillovers, that is, running a 2SLS using $Z_{ig}$ as an instrument for $D_{ig}$ without accounting for peer's assignment or treatment status. Taken at face value, these estimates suggest an ITT effect of about 2 percentage points and a local average effect of about 4 percentage points on voter turnout. However, in the presence of spillovers, these magnitudes do not generally have a clear causal interpretation. In fact, by Proposition \ref{prop:lapo_osnc}, given this assignment mechanism,
\begin{multline*}
\frac{\E[Y_{ig}|Z_{ig}=1]-\E[Y_{ig}|Z_{ig}=0]}{\E[D_{ig}|Z_{ig}=1]}=\\
\E[Y_{ig}(1,0)-Y_{ig}(0,0)|C_{ig}]-\E[Y_{ig}(0,1)-Y_{ig}(0,0)|C_{jg}]\P[Z_{jg}=1|Z_{ig}=0],
\end{multline*}
so the naive 2SLS estimand that ignores spillovers equals the difference between the direct and indirect LATEs, where the indirect LATE is rescaled by the conditional probability of treatment assignment. Therefore, the naive 2SLS estimand can be close to zero whenever direct and spillover effects have the same sign.

Column (2) in Table \ref{tab:results} shows the estimated direct and indirect ITT and LATE parameters based on Corollary \ref{coro:late_osnc}. The results reveal strong evidence of direct and indirect effects. On the one hand, ITT effects are positive and significant. ITT estimates are around 3 and 7 percentage points, respectively. While these magnitudes do not directly measure causal effects, as shown in Lemma \ref{lemma:direct_ITT}, under one-sided noncompliance ITT parameters are attenuated (rescaled) measures of local average effects, that is, $\E[Y_{ig}|Z_{ig}=1,Z_{jg}=0]-\E[Y_{ig}|Z_{ig}=0,Z_{jg}=0]=\E[Y_{ig}(1,0)|C_{ig}]\P[C_{ig}]$ and $\E[Y_{ig}|Z_{ig}=0,Z_{jg}=1]-\E[Y_{ig}|Z_{ig}=0,Z_{jg}=0]=\E[Y_{ig}(0,1)|C_{jg}]\P[C_{jg}]$. These magnitudes can be interpreted as the ``effect'' of offering the treatment (see the discussion on ITT and LATE in Section \ref{sec:ITT}).

On the other hand, 2SLS estimates reveal that the phone message increases voter turnout on compliers with untreated peers by about 7 percentage points, and turnout for untreated individuals with treated compliant peers by about 10 percentage points, both effects significant at the 1 percent level. Although the first-stage estimate suggests a strong instrument, I also calculate weak-instrument-robust AR 95\%-confidence intervals as a robustness check. These intervals are shown in parentheses in Table \ref{tab:results}. Reassuringly, the AR confidence intervals are slightly wider but very close to the large-sample-based confidence intervals, and lead to the same conclusions.

The finding that the estimated spillover effect is larger than the direct effect may seem surprising, as one may intuitively expect indirect effects to be weaker than direct ones. This comparison, however, must be done with care, as the estimated effects correspond to different subpopulations. More precisely, the direct effect is estimated for compliers, whereas the spillover effect is estimated for units with compliant peers, averaging over own compliance type. This is different than comparing the direct and spillover effects on the population of compliers. Note that the indirect LATE is:
\begin{align*}
\E[Y_{ig}(0,1)-Y_{ig}(0,0)|C_{jg}]&=\E[Y_{ig}(0,1)-Y_{ig}(0,0)|C_{ig},C_{jg}]\P[C_{ig}|C_{jg}]\\
&+\E[Y_{ig}(0,1)-Y_{ig}(0,0)|C^c_{ig},C_{jg}]\P[C^c_{ig}|C_{jg}]
\end{align*}
so it combines the effects on compliers and non-compliers, conditional on them having a compliant peer.

Finally, instrument validity can be assessed based on Proposition \ref{prop:spec}. In this application, because the outcome variable is binary, it is sufficient to run the following regression:
\begin{align}\label{eq:spec}
Y_{ig}(1-D_{ig})(1-D_{jg})=\gamma_0+\gamma_1 Z_{ig}+\gamma_2 Z_{jg}+u_{ig}
\end{align}
and test that $\gamma_1\le 0$ and $\gamma_2\le 0$. The results from this regression are shown in Table \ref{tab:spec}. As expected, both coefficients are negative and significant. 

\begin{table}
\begin{center}
\caption{Main Estimation Results\label{tab:results}}
\begin{tabular}{lcc}
\hline\hline
\multicolumn{1}{l}{}&\multicolumn{1}{c}{(1)}&\multicolumn{1}{c}{(2)}\tabularnewline
\hline
{\bfseries ITT}&&\tabularnewline
~~$Z_{ig}$&0.0178&0.0305\tabularnewline
~~&(0.0089)&(0.0114)\tabularnewline
~~&[0.0004 , 0.0352]&[0.0081 , 0.0529]\tabularnewline
~~$Z_{jg}$&&0.0459\tabularnewline
~~&&(0.0116)\tabularnewline
~~&&[0.0232 , 0.0686]\tabularnewline
\hline
{\bfseries 2SLS}&&\tabularnewline
~~$D_{ig}$&0.0394&0.0675\tabularnewline
~~&(0.0196)&(0.0252)\tabularnewline
~~&[0.0010 , 0.0778]&[0.0182 , 0.1168]\tabularnewline
~~&(0.0008 , 0.0782)&(0.0175 , 0.1208)\tabularnewline
~~$D_{jg}$&&0.1017\tabularnewline
~~&&(0.0255)\tabularnewline
~~&&[0.0517 , 0.1518]\tabularnewline
~~&&(0.0502 , 0.1570)\tabularnewline
\hline
{\bfseries }&&\tabularnewline
~~$N$&9,860&9,860\tabularnewline
~~Clusters&4,930&4,930\tabularnewline
\hline
\end{tabular}

\end{center}
\footnotesize \textbf{Notes}: estimated results from reduced-form regressions (``ITT'') and 2SLS regressions (``2SLS''). Column (1) shows the naive reduced-form and 2SLS estimates that ignore spillovers. Column (2) shows the ITT effects and local average direct and spillover effects obtained by 2SLS. Standard errors in parentheses. 95\%-confidence intervals in brackets are based on the large-sample normal approximation. 95\%-confidence intervals in parentheses are weak-instrument-robust AR confidence intervals. Estimation accounts for clustering at the household level.
\end{table}

\begin{table}
\begin{center}
\caption{Testing instrument validity\label{tab:spec}}
\begin{tabular}{lc}
\hline\hline
\multicolumn{1}{l}{}&\multicolumn{1}{c}{(1)}\tabularnewline
\hline
{\bfseries }&\tabularnewline
~~$Z_i$&$-0.0996$\tabularnewline
~~&($0.0094$)\tabularnewline
~~&[$-0.1181$ , $-0.0811$]\tabularnewline
~~$Z_j$&$-0.0893$\tabularnewline
~~&($0.0096$)\tabularnewline
~~&[$-0.1082$ , $-0.0704$]\tabularnewline
\hline
{\bfseries }&\tabularnewline
~~$N$&9,860\tabularnewline
~~Clusters&4,930\tabularnewline
\hline
\end{tabular}

\end{center}
\footnotesize \textbf{Notes}:  estimated results from Equation \eqref{eq:spec}. Standard errors in parentheses. 95\%-confidence intervals based on the large-sample normal approximation. Estimation accounts for clustering at the household level.
\end{table}


\section{Generalizations and Extensions}\label{sec:extensions}

\subsection{Conditional-on-observables IV}\label{sec:cia}

In many cases, the assumption that the instruments are as good as randomly assigned is more credible after conditioning on a set of covariates. This is particularly relevant when the variation in the instruments is not controlled by the researcher, but instead comes from a natural experiment. Furthermore, \citet{Abadie_2003_JoE} shows that covariates can be exploited to identify the characteristics of the compliant subpopulation.

In this section I generalize my results to the case in which quasi-random assignment of $(Z_{ig},Z_{jg})$ holds after conditioning on observable characteristics, following \citet{Abadie_2003_JoE}. Let $X_g=(X_{ig}',X_{jg}')'$ be a vector of observable characteristics for units $i$ and $j$ in group $g$. I introduce the following assumption.

\begin{assumption}[Conditional-on-observables IV]\label{assu:cia} $ $
\begin{enumerate}
\item Exclusion restriction: $Y_{ig}(d,d',z,z')=Y_{ig}(d,d')$ for all $(d,d',z,z')$,
\item Independence: for all $i$, $j\ne i$ and $g$, $((Y_{ig}(d,d'))_{(d,d')},(D_{ig}(z,z'))_{(z,z')})\indep (Z_{ig},Z_{jg})|X_g$,
\item Monotonicity: $\P[D_{ig}(1,1)\ge D_{ig}(1,0)\ge D_{ig}(0,1)\ge D_{ig}(0,0)|X_g]=1$,
\item One-sided noncompliance: $\P[D_{ig}(0,z')=0|X_g]=1$ for $z'=0,1$.
\end{enumerate}
\end{assumption}

Let $p_{zz'}(X_g)=\P[Z_{ig}=z,Z_{jg}=z'|X_g]$. Then we have the following result. 

\begin{proposition}[Identification conditional on observables]\label{prop:cia}
Under Assumption \ref{assu:cia},
\begin{align*}
\P[C_{ig}|X_g]&=\E[D_{ig}|Z_{ig}=1,Z_{jg}=0,X_g]\\
\P[GC_{ig}|X_g]&=\E[D_{ig}|Z_{ig}=1,Z_{jg}=1,X_g]-\E[D_{ig}|Z_{ig}=1,Z_{jg}=0,X_g] \\
\P[NT_{ig}|X_g]&=1-\P[GC_{ig}|X_g]-\P[C_{ig}|X_g]
\end{align*}
and for any (integrable) function $g(\cdot,\cdot)$,
\begin{align*}
\E[g(Y_{ig}(0,0),X_g)]&=\E\left[g(Y_{ig},X_g)\frac{(1-Z_{ig})(1-Z_{jg})}{p_{00}(X_g)}\right] \\
\E[g(Y_{ig}(1,0),X_g)|C_{ig}]\P[C_{ig}]&=\E\left[g(Y_{ig},X_g)D_{ig}\frac{Z_{ig}(1-Z_{ig})}{p_{10}(X_g)}\right] \\
\E[g(Y_{ig}(0,1),X_g)|C_{jg}]\P[C_{jg}]&=\E\left[g(Y_{ig},X_g)D_{jg}\frac{(1-Z_{ig})Z_{ig}}{p_{01}(X_g)}\right] \\
\E[g(Y_{ig}(0,0),X_g)|C_{ig}]\P[C_{ig}]&=\E\left[g(Y_{ig},X_g)\frac{(1-Z_{ig})(1-Z_{jg})}{p_{00}(X_g)}\right] -\E\left[g(Y_{ig},X_g)(1-D_{ig})\frac{Z_{ig}(1-Z_{jg})}{p_{10}(X_g)}\right] \\
\E[g(Y_{ig}(0,0),X_g)|C_{jg}]\P[C_{jg}]&=\E\left[g(Y_{ig},X_g)\frac{(1-Z_{ig})(1-Z_{jg})}{p_{00}(X_g)}\right] -\E\left[g(Y_{ig},X_g)(1-D_{jg})\frac{(1-Z_{ig})Z_{jg}}{p_{01}(X_g)}\right],
\end{align*}
whenever the required conditional probabilities $p_{zz'}(X_g)$ are positive. Furthermore, these equalities also hold conditional on $X_g$.
\end{proposition}

This result shows identification of functions of potential outcomes and covariates for compliers and for units with compliant peers. In particular, note that setting $g(y,x)=y$ recovers the result from Proposition \ref{prop:lapo_osnc}, which gives identification of local direct and spillover effects, both unconditionally or conditional on $X_g$. On the other hand, setting $g(y,x)=x$ shows that it is possible to identify the average characteristics of compliers and units with compliant peers. Hence, even if compliance type is unobservable, it is possible to characterize the distribution of observable characteristics for these subgroups. Estimation in this case can be based on reweighting methods after estimating the propensity score, as in \citet{Abadie_2003_JoE}.


\subsection{Multiple Units Per Group}\label{sec:sizes}

Without further assumptions, identification becomes increasingly harder as group size grows. The larger the group, the larger the set of compliance types, as units may respond in different ways to the different possible combinations of own and peers' treatment assignments, and it is generally not possible to pin down each unit's type. One-sided noncompliance is not enough to identify causal parameters when more than one unit is assigned to treatment since, as soon as more than one unit is assigned to treatment, it is not possible to distinguish between compliers and group compliers or between group compliers and never-takers.

Some studies addressed this problem by restricting the way in which potential treatment status depends on peers' assignments \citep[see e.g.][]{Kang-Imbens_2016_wp,Imai-Jiang-Malani_2020_JASA,DiTraglia-etal_2021}. In this section I provide an alternative assumption, \textit{independence of peers's types} (IPT), under which average potential outcomes do not depend on peers' compliance types. I then show that, under a generalization of the monotonicity assumption, average potential outcomes can be identified in the presence of two-sided noncompliance and spillovers in both outcomes and treatment statuses.

Suppose that each group $g$ has $n_g+1$ identically-distributed units, so that each unit in group $g$ has $n_g$ neighbors or peers. The vector of treatment statuses in each group is given by $\mathbf{D}_g=(D_{1g},\ldots,D_{n_g+1,g})$. For each unit $i$, $D_{j,ig}$ is the treatment indicator corresponding to unit $i$'s $j$-th neighbor, collected in the vector $\mathbf{D}_{(i)g}=(D_{1,ig},D_{2,ig},\ldots,D_{n_g,ig})$. This vector takes values $\mathbf{d}_g=(d_1,d_2,\ldots,d_{n_g})\in \mathcal{D}_g\subseteq \{0,1\}^{n_g}$. For a given realization of the treatment status $(d,\mathbf{d}_g)$, the potential outcome for unit $i$ in group $g$ is $Y_{ig}(d,\mathbf{d}_g)$ with observed outcome $Y_{ig}=Y_{ig}(D_{ig},\mathbf{D}_{(i)g})$. In what follows, $\mathbf{0}_ g$ and $\mathbf{1}_g$ will denote $n_g$-dimensional vectors of zeros and ones, respectively, and the analysis is conducted for a given group size $n_g$.

Let $\mathbf{Z}_{(i)g}$ be the vector of unit $i$'s peers' instruments, taking values $\mathbf{z}_g\in\{0,1\}^{n_g}$. To reduce the complexity of the model, I will assume that potential statuses and outcomes satisfy an exchangeability condition under which the identities of the treated peers do not matter, and thus the variables depend on the vectors $\mathbf{z}_g$ and $\mathbf{d}_g$, respectively, only through the sum of its elements. Under this condition, we have that $D_{ig}(z,\mathbf{z}_g)=D_{ig}(z,w_g)$ where $w_g=\mathbf{1}_g'\mathbf{z}_g$ and $Y_{ig}(d,\mathbf{d}_g)=Y_{ig}(d,s_g)$ where $s_g=\mathbf{1}_g'\mathbf{d}_g$.

The following assumption collects the required conditions for the upcoming results.

\begin{assumption}[Identification conditions for general $n_g$]\label{assu:general} $ $
\begin{enumerate}[(a)]
\item Exclusion restriction: $Y_{ig}(d,\mathbf{d}_g,\mathbf{z}_g)=Y_{ig}(d,\mathbf{d}_g,\mathbf{\tilde{z}}_g)$ for all $\mathbf{z}_g,\mathbf{\tilde{z}}_g$.
\item Independence: let $\mathbf{y}_{ig}=(Y_{ig}(d,\mathbf{d}_g))_{(d,\mathbf{d}_g)}$, $\mathbf{y}_{(i)g}=(\mathbf{y}_{jg})_{j\ne i}$, $\mathbf{\bar{d}}_{ig}=(D_{ig}(z,\mathbf{z}_g))_{(z,\mathbf{z}_g)}$ and $\mathbf{\bar{d}}_{(i)g}=(\mathbf{\bar{d}}_{jg})_{j\ne i}$. Then, $(\mathbf{y}_{ig},\mathbf{\bar{d}}_{ig},\mathbf{y}_{(i)g},\mathbf{\bar{d}}_{(i)g}) \indep (Z_{ig},\mathbf{Z}_{(i)g})$.
\item Exchangeability:
	\begin{enumerate}
	\item $D_{ig}(z,\mathbf{z}_g)=D_{ig}(z,w_g)$ where $w_g=\mathbf{1}_g'\mathbf{z}_g$.
	\item $Y_{ig}(d,\mathbf{d}_g)=Y_{ig}(d,s_g)$ where $s_g=\mathbf{1}_g'\mathbf{d}_g$.
	\end{enumerate}
\item Monotonicity:
	\begin{enumerate}[(i)]
	\item $D_{ig}(z,w_g)\ge D_{ig}(z,w_g')$ for $w_g\ge w_g'$ and $z=0,1$,
	\item $D_{ig}(1,0)\ge D_{ig}(0,n_g)$.
	\end{enumerate}
\end{enumerate}
\end{assumption}

Parts (a) and (b) in Assumption \ref{assu:general} generalize Assumptions \ref{assu:excl_restr} and \ref{assu:indep} to the case of general group sizes. Part (c) imposes exchangeability as discussed above. Part (d) generalizes the monotonicity assumption requiring that treatment take-up becomes more likely as the number of peers assigned to treatment increases, and that the effect of own assignment is stronger than the effect of peers' assignments.

Under monotonicity, we can define five compliance classes. First, always-takers, AT, are units with $D_{ig}(0,0)=1$ which implies $D_{ig}(z,w_g)=1$ for all $(z,w_g)$. Next, $w^*$-social compliers, SC($w^*$), are units for whom $D_{ig}(1,w_g)=1$ for all $w_g$, and for which there exists $0<w^*<n_g$ such that $D_{ig}(0,w_g)=1$ for all $w_g\ge w^*$. Thus, $w^*$-social compliers start receiving treatment as soon as $w^*$ of their peers are assigned to treatment. Compliers, C, are units with $D_{ig}(1,w_g)=1$ and $D_{ig}(0,w_g)=0$ for all $w_g$. Next, $w^*$-group compliers, GC($w^*$) have $D_{ig}(0,w_g)=0$ for all $w_g$ and there exists $0<w^*<n_g$ such that $D_{ig}(1,w_g)=1$ for all $w_g\ge w^*$. That is, $w^*$-group compliers need to be assigned to treatment and have at least $w^*$ peers assigned to treatment to actually receive the treatment. Finally, never-takers, NT, are units with $D_{ig}(z,w_g)=0$ for all $(z,w_g)$. Let $\xi_{ig}$ be a random variable indicating unit $i$'s compliance type, with $\xi_{ig}\in \Xi=\{\mathrm{NT},\mathrm{GC}(w^*),\mathrm{C},\mathrm{SC}(w^*),\mathrm{AT}|w^*=1,\ldots,n_g\}$ and $\boldsymbol{\xi}_{(i)g}$ the vector collecting $\xi_{jg}$ for $j\ne i$. As before, let the event $AT_{ig}=\{\xi_{ig}=\mathrm{AT}\}$, $C_{ig}=\{\xi_{ig}=\mathrm{C}\}$, $GC(\omega^*)=\{\xi_{ig}=\mathrm{GC}(\omega^*)\}$ and similarly for the other compliance types. Finally, let $W_{ig}=\sum_{j\ne i}Z_{jg}$ be the observed number of unit $i$' peers assigned to treatment. The following result discusses identification of the distribution of compliance types. 

\begin{proposition}\label{prop:distr_comp_g}
Under Assumption \ref{assu:general}, 
\begin{align*}
\P[AT_{ig}]&=\E[D_{ig}|Z_{ig}=0,W_{ig}=0] \\
\P[SC_{ig}(w^*)]&=\E[D_{ig}|Z_{ig}=0,W_{ig}=w^*]-\E[D_{ig}|Z_{ig}=0,W_{ig}=w^*-1],\quad 1<w^*<n_g \\
\P[C_{ig}]&=\E[D_{ig}|Z_{ig}=1,W_{ig}=0]-\E[D_{ig}|Z_{ig}=0,W_{ig}=n_g] \\
\P[GC_{ig}(w^*)]&=\E[D_{ig}|Z_{ig}=1,W_{ig}=w^*]-\E[D_{ig}|Z_{ig}=1,W_{ig}=w^*-1],\quad 1<w^*<n_g \\
\P[NT_{ig}]&=\E[1-D_{ig}|Z_{ig}=1,W_{ig}=n_g].
\end{align*}
\end{proposition}

The following assumption restricts the amount of heterogeneity in potential outcomes by ensuring that potential outcomes are independent from peers' compliance types, conditional on own type.

\begin{assumption}[Independence of peers' types]\label{assu:ipt}
Let $\mathbf{y}_{ig}=(Y_{ig}(d,\mathbf{d}_g))_{(d,\mathbf{d}_g)}$. Potential outcomes are independent of peers' types: $\mathbf{y}_{ig}\indep \boldsymbol{\xi}_{(i)g} \,|\, \xi_{ig}$.
\end{assumption}

Intuitively, IPT states that own type summarizes all the heterogeneity in potential outcome distributions. For example, this assumption may hold when groups are randomly formed, so that types are independent, or when types are perfectly correlated (i.e. units match with other units of the same type), so that conditional on own type, peers' types are constant.

The following result shows which average potential outcomes are identified under these assumptions.

\begin{proposition}[Identification under IPT]\label{prop:lapo_ipt1}
Under Assumptions \ref{assu:general} and \ref{assu:ipt}, if $\P[D_{ig}=d,Z_{ig}=z,S_{ig}=s,W_{ig}=w]>0$, then
$\E[Y_{ig}|D_{ig}=d,Z_{ig}=z,S_{ig}=s,W_{ig}=w]=\E[Y_{ig}(d,s)|D_{ig}(z,w)=d]$. In particular, if $\P[D_{ig}=d,Z_{ig}=z,S_{ig}=s,W_{ig}=w]>0$ for all $(d,z,s,w)$, then $\E[Y_{ig}(1,s)|\xi_{ig}]$ is identified for all $s$ and $\xi_{ig}\ne \mathrm{NT}$, and $\E[Y_{ig}(0,s)|\xi_{ig}]$ is identified for all $s$ and $\xi_{ig}\ne \mathrm{AT}$.
\end{proposition}

In particular, this result implies that all the average potential outcomes for compliers $\E[Y_{ig}(d,s_g)|C_{ig}]$ are identified. See the proof in the supplemental appendix for further details.


\section{Conclusion}\label{sec:conclusion}

This paper proposed a potential outcomes framework to analyze identification and estimation of causal spillover effects using instrumental variables. The findings in this paper highlight the challenges of analyzing spillover effects with imperfect compliance and provide practical guidance on how to address them. First, when groups consist of pairs (such as couples, roommates, siblings), local average effects can be identified under one-sided noncompliance using 2SLS methods. One advantage of this approach is that one-sided noncompliance can be straightforwardly verified in practice. While 2SLS methods do not work in general under two-sided noncompliance or when units have multiple peers, the independence of peers' types assumption introduced in Section \ref{sec:sizes} provides an alternative restriction on effect heterogeneity that permits identification of causally interpretable parameters. Finally, Section \ref{sec:cia} generalizes the results to cases in which the assumption of as-if random assignment of the instruments holds after conditioning on a set of covariates, which allows the researcher to apply the results in this paper in more general settings such as natural experiments.

\newpage

\bibliographystyle{jasa}
\bibliography{references}

\newpage

\begin{appendices}
\section{Additional Identification Results}\label{app:addresults}

\subsection{Indirect ITT Effects}

The following result characterizes the indirect ITT effect.

\begin{lemma}\label{lemma:indirect_ITT}
Under Assumptions \ref{assu:excl_restr}-\ref{assu:monot},
\begin{align*}
\E[Y_{ig}|Z_{ig}=0,Z_{jg}=1]-\E[Y_{ig}|&Z_{ig}=0,Z_{jg}=0]=\\
&\E[Y_{ig}(1,0)-Y_{ig}(0,0)|\{SC_{ig}\}\times \{GC_{jg},NT_{jg}\}]\\
&\qquad \times \P[\{SC_{ig}\}\times \{GC_{jg},NT_{jg}\}]\\
+&\E[Y_{ig}(1,1)-Y_{ig}(0,0)|\{SC_{ig}\}\times\{SC_{jg},C_{jg}\}]\\
&\qquad \times \P[\{SC_{ig}\}\times\{SC_{jg},C_{jg}\}]\\
+&\E[Y_{ig}(1,1)-Y_{ig}(0,1)|SC_{ig},AT_{jg}]\\
&\qquad\times \P[SC_{ig},AT_{jg}]\\
+&\E[Y_{ig}(0,1)-Y_{ig}(0,0)|\{C_{ig},GC_{ig},NT_{ig}\}\times\{SC_{jg},C_{jg}\}]\\
&\qquad \times \P[\{C_{ig},GC_{ig},NT_{ig}\}\times\{SC_{jg},C_{jg}\}]\\
+&\E[Y_{ig}(1,1)-Y_{ig}(1,0)|\{AT_{ig}\}\times\{SC_{jg},C_{jg}\}] \\
&\qquad \times \P[\{AT_{ig}\}\times\{SC_{jg},C_{jg}\}].
\end{align*}
\end{lemma}


\subsection{Total ITT Effects}

The following result characterizes the total ITT effect.

\begin{lemma}\label{lemma:total_ITT}
Under Assumptions \ref{assu:excl_restr}-\ref{assu:monot}
\begin{align*}
\E[Y_{ig}|Z_{ig}=1,Z_{jg}=1]-\E[Y_{ig}|&Z_{ig}=0,Z_{jg}=0]=\\
&\E[Y_{ig}(1,0)-Y_{ig}(0,0)|\{SC_{ig},C_{ig},GC_{ig}\}\times \{NT_{jg}\}]\\
&\qquad \times \P[\{SC_{ig},C_{ig},GC_{ig}\}\times \{NT_{jg}\}]\\
+&\E[Y_{ig}(1,1)-Y_{ig}(1,0)|\{AT_{ig}\}\times\{SC_{jg},C_{jg},GC_{jg}\}]\\
&\qquad \times \P[\{AT_{ig}\}\times\{SC_{jg},C_{jg},GC_{jg}\}]\\
+&\E[Y_{ig}(0,1)-Y_{ig}(0,0)|\{NT_{ig}\},\{SC_{ig},C_{jg},GC_{jg}\}]\\
&\qquad\times \P[\{NT_{ig}\},\{SC_{ig},C_{jg},GC_{jg}\}]\\
+&\E[Y_{ig}(1,1)-Y_{ig}(0,1)|\{SC_{ig},C_{ig},GC_{ig}\}\times\{AT_{jg}\}]\\
&\qquad \times \P[\{SC_{ig},C_{ig},GC_{ig}\}\times\{AT_{jg}\}]\\
+&\E[Y_{ig}(1,1)-Y_{ig}(0,0)|\{SC_{ig},C_{ig},GC_{ig}\}\times\{SC_{jg},C_{jg},GC_{jg}\}] \\
&\qquad \times \P[\{SC_{ig},C_{ig},GC_{ig}\}\times\{SC_{jg},C_{jg},GC_{jg}\}].
\end{align*}
\end{lemma}


\subsection{Identification Under Monotonicity}\label{app:identif_monot}

In the absence of spillovers, \citet{Imbens-Rubin_1997_ReStud} show that different combinations of $D_{ig}$ and $Z_{ig}$ can be exploited to identify average potential outcomes for compliers. The intuition behind this approach is that a unit with $D_{ig}=1$ and $Z_{ig}=0$ is necessarily an always-taker, whereas a unit with $D_{ig}=1$ and $Z_{ig}=1$ can be an always-taker or a complier, and hence the combination of these two cases identifies $\E[Y_{ig(1)}|C_{ig}]$ (and an analogous argument implies identification of $\E[Y_{ig}(0)|C_{ig}]$). To see why does approach does not work in the presence of spillovers, notice that a unit with $D_{ig}=1,D_{jg}=1,Z_{ig}=0,Z_{jg}=0$ is an always-taker with an always-taker peer. However, a unit with $D_{ig}=1,D_{jg}=1,Z_{ig}=1,Z_{jg}=0$ could be an always-taker, a social complier or a complier, and her peer could be an always-taker or a social complier, and it is not possible to disentangle each unit's compliance type.

Table \ref{tab:identif_monot} shows what can be identified under monotonicity without further restrictions by exploiting the variation in $(D_{ig},D_{jg},Z_{ig},Z_{jg})$. For instance, the first row in the table indicates that:
\[\E[Y_{ig}|D_{ig}=1,D_{jg}=1,Z_{ig}=0,Z_{jg}=0]=\E[Y_{ig}(1,1)|AT_{ig},AT_{jg}].\]

The table shows that, without further assumptions, is it not possible to point identify average potential outcomes for specific compliance types, with the exception of $\E[Y_{ig}(1,1)|AT_{ig},AT_{jg}]$ and $\E[Y_{ig}(0,0)|NT_{ig},NT_{jg}]$.

On the other hand, Table \ref{tab:identif_monot_OSN} illustrates why one-sided noncompliance is enough to obtain identification of causal parameters. Importantly, under one-sided noncompliance, the event $C_{ig}\cup GC_{ig}\cup NT_{ig}$ covers the whole sample space, and therefore, for example, $\E[Y_{ig}|D_{ig}=1,D_{jg}=0,Z_{ig}=1,Z_{jg}=0]=\E[Y_{ig}(1,0)|\{C_{ig}\}\times \{C_{ig},GC_{ig},NT_{ig}\}]=\E[Y_{ig}(1,0)|C_{ig}]$.

Finally, Table \ref{tab:identif_monot_AT} illustrates that this strategy does not work when ruling out always-takers only.

\begin{table}
\caption{System of equations under monotonicity}\label{tab:identif_monot}
\begin{center}
\begin{tabular}{ccccccc}
$D_{ig}$ & $D_{jg} $ & $Z_{ig}$ & $Z_{jg}$ & $Y_{ig}(d,d')$ & $\xi_{ig}$ & $\xi_{jg}$ \\ \hline \hline
1 & 1 & 0 & 0 & $Y_{ig}(1,1)$ & AT & AT \\
1 & 1 & 0 & 1 & $Y_{ig}(1,1)$ & AT,SC & AT,SC,C\\
1 & 1 & 1 & 0 & $Y_{ig}(1,1)$ & AT,SC,C & AT,SC\\
1 & 1 & 1 & 1 & $Y_{ig}(1,1)$ & AT,SC,C,GC & AT,SC,C,GC\\ \hline
0 & 0 & 1 & 1 & $Y_{ig}(0,0)$ & NT & NT \\
0 & 0 & 1 & 0 & $Y_{ig}(0,0)$ & GC,NT & C,GC,NT \\
0 & 0 & 0 & 1 & $Y_{ig}(0,0)$ & C,GC,NT & GC,NT\\
0 & 0 & 0 & 0 & $Y_{ig}(0,0)$ & SC,C,GC,NT & SC,C,GC,NT\\ \hline
1 & 0 & 0 & 0 & $Y_{ig}(1,0)$ & AT & SC,C,GC,NT \\
1 & 0 & 0 & 1 & $Y_{ig}(1,0)$ & AT,SC & GC,NT \\
1 & 0 & 1 & 0 & $Y_{ig}(1,0)$ & AT,SC,C & C,GC,NT \\
1 & 0 & 1 & 1 & $Y_{ig}(1,0)$ & AT,SC,C,GC & NT \\ \hline
0 & 1 & 1 & 1 & $Y_{ig}(0,1)$ & NT & AT,SC,C,GC\\
0 & 1 & 1 & 0 & $Y_{ig}(0,1)$ & GC,NT & AT,SC\\
0 & 1 & 0 & 1 & $Y_{ig}(0,1)$ & C,GC,NT & AT,SC,C \\
0 & 1 & 0 & 0 & $Y_{ig}(0,1)$ & SC,C,GC,NT & AT\\  \hline
\end{tabular}
\end{center}
\end{table}

\begin{table}
\caption{System of equations under monotonicity and OSN}\label{tab:identif_monot_OSN}
\begin{center}
\begin{tabular}{ccccccc}
$D_{ig}$ & $D_{jg} $ & $Z_{ig}$ & $Z_{jg}$ & $Y_{ig}(d,d')$ & $\xi_{ig}$ & $\xi_{jg}$ \\ \hline \hline
1 & 1 & 1 & 1 & $Y_{ig}(1,1)$ & C,GC & C,GC\\ \hline
0 & 0 & 1 & 1 & $Y_{ig}(0,0)$ & NT & NT \\
0 & 0 & 1 & 0 & $Y_{ig}(0,0)$ & GC,NT & C,GC,NT \\
0 & 0 & 0 & 1 & $Y_{ig}(0,0)$ & C,GC,NT & GC,NT\\
0 & 0 & 0 & 0 & $Y_{ig}(0,0)$ & C,GC,NT & C,GC,NT\\ \hline
1 & 0 & 1 & 0 & $Y_{ig}(1,0)$ & C & C,GC,NT \\
1 & 0 & 1 & 1 & $Y_{ig}(1,0)$ & C,GC & NT \\ \hline
0 & 1 & 1 & 1 & $Y_{ig}(0,1)$ & NT & C,GC\\
0 & 1 & 0 & 1 & $Y_{ig}(0,1)$ & C,GC,NT & C \\ \hline
\end{tabular}
\end{center}
\end{table}

\begin{table}
\caption{System of equations under monotonicity and no AT}\label{tab:identif_monot_AT}
\begin{center}
\begin{tabular}{ccccccc}
$D_{ig}$ & $D_{jg} $ & $Z_{ig}$ & $Z_{jg}$ & $Y_{ig}(d,d')$ & $\xi_{ig}$ & $\xi_{jg}$ \\ \hline \hline
1 & 1 & 0 & 1 & $Y_{ig}(1,1)$ & SC & SC,C\\
1 & 1 & 1 & 0 & $Y_{ig}(1,1)$ & SC,C & SC\\
1 & 1 & 1 & 1 & $Y_{ig}(1,1)$ & SC,C,GC & SC,C,GC\\ \hline
0 & 0 & 1 & 1 & $Y_{ig}(0,0)$ & NT & NT \\
0 & 0 & 1 & 0 & $Y_{ig}(0,0)$ & GC,NT & C,GC,NT \\
0 & 0 & 0 & 1 & $Y_{ig}(0,0)$ & C,GC,NT & GC,NT\\
0 & 0 & 0 & 0 & $Y_{ig}(0,0)$ & SC,C,GC,NT & SC,C,GC,NT\\ \hline
1 & 0 & 0 & 1 & $Y_{ig}(1,0)$ & SC & GC,NT \\
1 & 0 & 1 & 0 & $Y_{ig}(1,0)$ & SC,C & C,GC,NT \\
1 & 0 & 1 & 1 & $Y_{ig}(1,0)$ & SC,C,GC & NT \\ \hline
0 & 1 & 1 & 1 & $Y_{ig}(0,1)$ & NT & SC,C,GC\\
0 & 1 & 1 & 0 & $Y_{ig}(0,1)$ & GC,NT & SC\\
0 & 1 & 0 & 1 & $Y_{ig}(0,1)$ & C,GC,NT & SC,C \\ \hline
\end{tabular}
\end{center}
\end{table}


\subsection{Multiple Treatment Levels}\label{app:multi}

The identification results in the paper can be extended to the case of multi-level treatments. To adapt the notation to this case, suppose that $D_{ig}\in\mathcal{D}=\{0,1,\ldots,K\}$. The potential outcome is $Y_{ig}(k,k')$ (which implicitly imposes the exclusion restriction) where $k,k'\in\{0,1,\ldots,K\}$ indicate different treatment levels. The observed outcome is:
\[Y_{ig}=\sum_{k\in\mathcal{D}}\sum_{k'\in\mathcal{D}}Y_{ig}(k,k')\1(D_{ig}=k)\1(D_{jg}=k').\]
Suppose that the instrument $Z_{ig}$ takes as many values as the treatment, that is, $Z_{ig}\in\mathcal{Z}$ where $\mathcal{Z}=\mathcal{D}$. For example, $Z_{ig}$ could indicate random assignment into different treatment levels, and $D_{ig}$ indicates whether unit $i$ actually receives the assigned treatment level. The potential treatment status is $D_{ig}(k,k')$ where $k,k'\in\mathcal{Z}$. Each unit $i$ can have up to $(K+1)^2$ different potential treatment statuses given by own and peer's treatment assignment.

I will assume that non-compliance is one-sided, and that units cannot switch between non-control treatment statuses. For example, in \citet{Foos-DeRooij_2017_AJPS}, the treatment assignment consists of three treatment levels: control, low-intensity treatment and high-intensity treatment. In this setting, units may refuse the treatment they are assigned, but units assigned to the low-intensity treatment cannot receive the high-intensity treatment and vice versa. 

\begin{proposition}\label{prop:multi}
Suppose that, in addition to Assumption \ref{assu:indep}, $\mathcal{D}=\mathcal{Z}=\{0,1,2,\ldots,K\}$, and:
\begin{enumerate}[(a)]
\item $D_{ig}(0,k')=0$ for all $k'\in\mathcal{Z}$,
\item $D_{ig}(k,k')\in\{0,k\}$ for all $k,k'\in\mathcal{Z}$.
\end{enumerate}
Then, for any $k$, $k'\in \{0,1,\ldots,K\}$ such that $\E[\1(D_{ig}(k,0)=k)]>0$ and $\E[\1(D_{jg}(0,k')=k')]>0$,
\begin{align*}
\E[\1(D_{ig}=k)|Z_{ig}=k,Z_{jg}=0]&=\E[\1(D_{ig}(k,0)=k)]\\
\frac{\E[Y_{ig}|Z_{ig}=k,Z_{jg}=0]-\E[Y_{ig}|Z_{ig}=0,Z_{jg}=0]}{\E[\1(D_{ig}=k)|Z_{ig}=k,Z_{jg}=0]}&=\E[Y_{ig}(k,0)-Y_{ig}(0,0)|D_{ig}(k,0)=k]\\
\frac{\E[Y_{ig}|Z_{ig}=0,Z_{jg}=k']-\E[Y_{ig}|Z_{ig}=0,Z_{jg}=0]}{\E[\1(D_{jg}=k')|Z_{ig}=0,Z_{jg}=k']}&=\E[Y_{ig}(0,k')-Y_{ig}(0,0)|D_{jg}(k',0)=k'].
\end{align*}
\end{proposition}

Condition (a) in Proposition \ref{prop:multi} implies that units who are assigned to the control condition remain untreated, and Condition (b) states that a unit who is offered treatment level $k$ can either receive that treatment level or remain untreated. This result shows how to identify the proportion of units who comply with treatment level $k$, $\E[\1(D_{ig}(k,0)=k)]$, the average direct effect on units who comply with treatment level $k$, $\E[Y_{ig}(k,0)-Y_{ig}(0,0)|D_{ig}(k,0)=k]$, and the average spillover effect on units whose peer complies with treatment level $k'$.

\section{Further Details on Estimation and Inference}\label{app:estimation}

All the parameters of interest in Section \ref{sec:identif} can be recovered by estimating expectations using sample means. More precisely, let
\[\1^\mathbf{z}_{ig}=\begin{bmatrix}
\1(Z_{ig}=0,Z_{jg}=0)\\
\1(Z_{ig}=1,Z_{jg}=0)\\
\1(Z_{ig}=0,Z_{jg}=1)\\
\1(Z_{ig}=1,Z_{jg}=1)
\end{bmatrix}\]
and let $\mathbf{H}(\cdot)$ be a vector-valued function whose exact shape depends on the parameters to be estimated, as illustrated below. Then the goal is to estimate:
\[\boldsymbol{\mu}=\E \begin{bmatrix}
\1^\mathbf{z}_{ig} \\
\mathbf{H}(Y_{ig},D_{ig},D_{jg}) \otimes \1^\mathbf{z}_{ig}
\end{bmatrix}\]
where the first four elements correspond to the assignment probabilities $\P[Z_{ig}=z,Z_{jg}=z']$ and the remaining elements corresponds to estimands of the form $\E[Y_{ig}\1(D_{ig}=d,D_{jg}=d')\1(Z_{ig}=z,Z_{jg}=z')]$. The most general choice of $\mathbf{H}$ in this setup is the following:
\[\mathbf{H}(Y_{ig},D_{ig},D_{jg})=\begin{bmatrix}
\1(D_{ig}=0,D_{jg}=0)\\
\vdots\\
\1(D_{ig}=1,D_{jg}=1) \\
Y_{ig}\1(D_{ig}=0,D_{jg}=0)\\
\vdots\\
Y_{ig}\1(D_{ig}=1,D_{jg}=1)
\end{bmatrix}\]
which is a vector of dimension equal to eight that can be used to estimate all the first-stage estimands $\E[\1(D_{ig}=d,D_{jg}=d')|Z_{ig}=z,Z_{jg}=z']$ and average outcomes $\E[Y_{ig}\1(D_{ig}=d,D_{jg}=d')|Z_{ig}=z,Z_{jg}=z']$. In this general case, the total number of equations to be estimated is 36: four probabilities $\P[Z_{ig}=z,Z_{jg}=z']$ plus the four indicators $\1(Z_{ig}=z,Z_{jg}=z')$ times each of the eight elements in $\mathbf{H}(\cdot)$. The dimension of $\mathbf{H}(\cdot)$ can be reduced, for example, by focusing on ITT parameters $\E[Y_{ig}|Z_{ig}=z,Z_{jg}=z']$, which corresponds to:
\[\mathbf{H}(Y_{ig},D_{ig},D_{jg})=Y_{ig},\]
or by imposing the assumptions described in previous sections. For instance, under one-sided noncompliance, the parameters in Corollaries \ref{coro:late_osnc} and \ref{coro:het_osnc} can be estimated by defining:
\[\1^\mathbf{z}_{ig}=\begin{bmatrix}
\1(Z_{ig}=0,Z_{jg}=0)\\
\1(Z_{ig}=1,Z_{jg}=0)\\
\1(Z_{ig}=0,Z_{jg}=1)
\end{bmatrix}\]
and
\[\mathbf{H}(Y_{ig},D_{ig},D_{jg})=\begin{bmatrix}
D_{ig}\\
Y_{ig} \\
Y_{ig}(1-D_{ig})\\
Y_{ig}(1-D_{jg})
\end{bmatrix}.\]

Regardless of the choice of $\1^\mathbf{z}_{ig}$ and $\mathbf{H}(\cdot)$, the dimension of the vector of parameters to be estimated is fixed (and at most 36). Consider the following sample mean estimator:
\[\boldsymbol{\hat{\mu}}=\frac{1}{G}\sum_{g=1}^G W_g\]
where 
\[W_g= \begin{bmatrix}
(\1^\mathbf{z}_{1g}+\1^\mathbf{z}_{2g})/2\\
(\mathbf{H}(Y_{1g},D_{1g},D_{2g}) \otimes \1^\mathbf{z}_{1g} + \mathbf{H}(Y_{2g},D_{2g},D_{1g}) \otimes \1^\mathbf{z}_{2g})/2
\end{bmatrix}.\]

It is straightforward to see that under assumption \ref{assu:sampling}, $\hat{\boldsymbol{\mu}}$ is consistent for $\boldsymbol{\mu}$ and converges in distribution to a normal random variable after centering and rescaling as $G\to\infty$:
\[\sqrt{G}(\boldsymbol{\hat{\mu}}-\boldsymbol{\mu}) \overset{\mathcal{D}}{\to}\mathcal{N}(\mathbf{0},\boldsymbol{\Sigma})\]
where $\boldsymbol{\Sigma}=\E[(W_g-\boldsymbol{\mu})(W_g-\boldsymbol{\mu})']$, and where the limiting variance can be consistently estimated by:
\[\boldsymbol{\hat{\Sigma}}=\frac{1}{G}\sum_g (W_g-\boldsymbol{\hat{\mu}})(W_g-\boldsymbol{\hat{\mu}})'.\]
Finally, once $\boldsymbol{\hat{\mu}}$ has been estimated, the treatment effects of interest can be estimated as (possibly nonlinear) transformations of $\boldsymbol{\hat{\mu}}$, and their variance estimated using the delta method. 


\section{Further Details on AR confidence intervals}\label{app:AR}

In this section I outline the procedure to construct weak-instrument-robust confidence intervals for the direct effect on compliers (the procedure for the spillover effect is analogous). Given some hypothesized value $\beta_1^*$ for this effect, consider the following regression on the subsample of units with untreated peers:
\[Y_{ig}-\beta_1^*D_{ig}=\theta_0+\theta_1Z_{ig}+\epsilon_{ig}.\]
Since $Z_{ig}$ is binary, $\hat{\theta}_1=\hat{\gamma}_1-\beta_1^*\bar{D}_{10}$ where $\bar{D}_{10}=\sum_{g,i}D_{ig}Z_{ig}(1-Z_{jg})/\sum_{g,i}Z_{ig}(1-Z_{jg})$ is the first-stage estimate and $\hat{\gamma}_1$ is the reduced-form estimate $\sum_{g,i}Y_{ig}Z_{ig}(1-Z_{jg})/\sum_{g,i}Z_{ig}(1-Z_{jg})-\sum_{g,i}Y_{ig}(1-Z_{ig})(1-Z_{jg})/\sum_{g,i}(1-Z_{ig})(1-Z_{jg})$. By previous results, as $G\to\infty$, $\hat{\theta}_1\to_\P\left(\E[Y_{ig}(1,0)-Y_{ig}(0,0)|C_{ig}]-\beta_1^*\right)\P[C_{ig}]$ and thus under the null hypothesis that $\E[Y_{ig}(1,0)-Y_{ig}(0,0)|C_{ig}]=\beta_1^*$, $\theta_1=0$ and:
\[\frac{\hat{\theta}_1^2}{V(\hat{\theta}_1)}\to_\mathcal{D}\chi^2_1\]
where $\chi^2_1$ is the chi-squared distribution and $V(\hat{\theta}_1)$ is the variance of $\hat{\theta}_1$. Noting that both $\hat{\theta}_1$ and its variance depend on $\beta_1^*$, an AR confidence interval for $\E[Y_{ig}(1,0)-Y_{ig}(0,0)|C_{ig}]$ is given by:
\[\{\beta_1^*:\hat{\theta}_1^2\le V(\hat{\theta}_1)\chi^2_{1,1-\alpha}\}.\]
Finally, since $\hat{\theta}_1=\hat{\gamma}_1-\beta_1^*\bar{D}_{10}$, $V(\hat{\theta}_1)=V(\hat{\gamma}_1)+(\beta_1^*)^2V(\bar{D}_{10})-2\beta_1^*Cov(\hat{\gamma}_1,\bar{D}_{10})$. Therefore, the AR confidence interval can be obtained by solving the inequality:
\[(\beta_1^*)^2\left[\bar{D}_{10}^2-\chi^2_{1,1-\alpha}V(\bar{D}_{10})\right]+2\beta_1^*\left[\chi^2_{1,1-\alpha}Cov(\hat{\gamma}_1,\bar{D}_{10})-\hat{\gamma}_1\bar{D}_{10}\right]+\hat{\gamma}_1^2-\chi^2_{1,1-\alpha}V(\hat{\gamma}_1)\le 0\]
which depends only on the vector of estimated coefficients, their variance matrix and a quantile from the $\chi^2_1$ distribution. In particular, notice that this quadratic function is strictly convex whenever $\bar{D}_{10}^2/V(\bar{D}_{10})>\chi^2_{1,1-\alpha}$, which holds when the null hypothesis the first-stage coefficient is zero is rejected at the $\alpha$ level. In this case, the AR confidence interval is bounded and convex.


\section{Additional Empirical Results}\label{app:multi_emp}

The experiment conducted by \citet{Foos-DeRooij_2017_AJPS} included two different treatment levels. More precisely, the sample of 5,190 two-voter households with landline numbers were stratified into three blocks based on their last recorded party preference (Labor party supporter, rival party supported, unattached) and randomly assigned to one of three treatment arms:
\begin{itemize}
\item High-intensity treatment: the telephone message had a strong partisan tone, explicitly mentioning the Labour party and policies, taking an antagonistic stance toward the main rival party.
\item Low-intensity treatment: the telephone message avoided statements about party competition and did not mention the candidate's affiliation nor the rival party.
\item Control: did not receive any form of contact from the campaign.
\end{itemize}
Finally, within the households assigned to the low- or high-intensity treatment arms, only one household member was randomly selected to receive the telephone message. 

In this section, I apply the results from Proposition \ref{prop:multi} to analyze the effect of each treatment arm by separately comparing households exposed to each treatment intensity to the control households. The empirical results are shown in Table \ref{tab:results_multi}. The estimates suggest that both treatment arms had very similar effects. 

\begin{table}
\begin{center}
\caption{Estimation Results with Multiple Treatments\label{tab:results_multi}}
\begin{tabular}{lcc}
\hline\hline
\multicolumn{1}{l}{}&\multicolumn{1}{c}{Low-intensity treatment}&\multicolumn{1}{c}{High-intensity treatment}\tabularnewline
\hline
{\bfseries ITT}&&\tabularnewline
~~$Z_{ig}$&0.0300&0.0310\tabularnewline
~~&(0.0145)&(0.0146)\tabularnewline
~~&[0.0016 , 0.0584]&[0.0023 , 0.0597]\tabularnewline
~~$Z_{jg}$&0.0485&0.0433\tabularnewline
~~&(0.0148)&(0.0149)\tabularnewline
~~&[0.0195 , 0.0775]&[0.0142 , 0.0724]\tabularnewline
\hline
{\bfseries 2SLS}&&\tabularnewline
~~$D_{ig}$&0.0662&0.0688\tabularnewline
~~&(0.0318)&(0.0323)\tabularnewline
~~&[0.0039 , 0.1285]&[0.0055 , 0.1322]\tabularnewline
~~$D_{jg}$&0.1070&0.0962\tabularnewline
~~&(0.0324)&(0.0328)\tabularnewline
~~&[0.0435 , 0.1706]&[0.0318 , 0.1606]\tabularnewline
\hline
{\bfseries }&&\tabularnewline
~~$N$&7,750&7,696\tabularnewline
~~Clusters&3,875&3,848\tabularnewline
\hline
\end{tabular}

\end{center}
\footnotesize \textbf{Notes}: estimated results from reduced-form regressions (``ITT'') and 2SLS regressions (``2SLS'') separately for each treatment arm against the control households. The first column shows the ITT and 2SLS results for the low-intensity treatment arm. The second column shows the ITT and 2SLS results for the high-intensity treatment arm. Standard errors in parentheses. 95\%-confidence intervals are based on the large-sample normal approximation. Estimation accounts for clustering at the household level.
\end{table}


\section{Proofs of Main Results}

\subsection{Proof of Proposition \ref{prop:comp_distr}}

By Assumption \ref{assu:indep}, $\E[D_{ig}|Z_{ig}=z,Z_{jg}=z']=\E[D_{ig}(z,z')]$. Thus, under monotonicity (Assumption \ref{assu:monot}),
\begin{align*}
\E[D_{ig}|Z_{ig}=0,Z_{ig}=0]&=\E[D_{ig}(0,0)]=\P[AT_{ig}] \\
\E[D_{ig}|Z_{ig}=0,Z_{ig}=1]&=\E[D_{ig}(0,1)]=\P[AT_{ig}]+\P[SC_{ig}] \\
\E[D_{ig}|Z_{ig}=1,Z_{ig}=0]&=\E[D_{ig}(1,0)]=\P[AT_{ig}]+\P[SC_{ig}]+\P[C_{ig}] \\
\E[D_{ig}|Z_{ig}=1,Z_{ig}=1]&=\E[D_{ig}(1,1)]=\P[AT_{ig}]+\P[SC_{ig}]+\P[C_{ig}]+\P[GC_{ig}]
\end{align*}
and by solving the system it follows that:
\begin{align*}
\P[AT_{ig}] &=\E[D_{ig}|Z_{ig}=0,Z_{ig}=0] \\
\P[SC_{ig}]&=\E[D_{ig}|Z_{ig}=0,Z_{ig}=1]-\E[D_{ig}|Z_{ig}=0,Z_{ig}=0]\\
\P[C_{ig}]&=\E[D_{ig}|Z_{ig}=1,Z_{ig}=0]-\E[D_{ig}|Z_{ig}=0,Z_{ig}=1]\\
\P[GC_{ig}]&=\E[D_{ig}|Z_{ig}=1,Z_{ig}=1]-\E[D_{ig}|Z_{ig}=1,Z_{ig}=0]
\end{align*}
and by monotonicity $\P[NT_{ig}]=1-\P[AT_{ig}]-\P[SC_{ig}]-\P[C_{ig}]-\P[GC_{ig}]$. Finally,
\begin{align*}
\E[D_{ig}D_{jg}|Z_{ig}=0,Z_{ig}=0]&=\E[D_{ig}(0,0)D_{jg}(0,0)]=\P[AT_{ig},AT_{jg}]\\
\E[(1-D_{ig})(1-D_{jg})|Z_{ig}=1,Z_{ig}=1]&=\E[(1-D_{ig}(1,1))(1-D_{jg}(1,1))]=\P[NT_{ig},NT_{jg}].
\end{align*}
See Tables \ref{tab:seq1} and \ref{tab:seq2} for the whole system of equations. $\square$

\begin{table}
\caption{System of equations}\label{tab:seq1}
\begin{center}
\begin{tabular}{ccccc}
$D_{ig}$ & $D_{jg} $ & $Z_{ig}$ & $Z_{jg}$ & Probabilities \\ \hline \hline
1 & 1 & 0 & 0 & $p_{AA}$ \\
1 & 1 & 0 & 1 & $p_{AA}+p_{AS}+p_{AC}+p_{SA}+p_{SS}+p_{SC}$ \\
1 & 1 & 1 & 0 & $p_{AA}+p_{AS}+p_{AC}+p_{SA}+p_{SS}+p_{SC}$ \\
1 & 1 & 1 & 1 & $1-2p_{N}+p_{NN}$ \\ \hline
0 & 0 & 1 & 1 & $p_{NN}$ \\
0 & 0 & 1 & 0 & $p_{GC}+p_{GG}+p_{GN}+p_{NC}+p_{NG}+p_{NN}$ \\
0 & 0 & 0 & 1 & $p_{GC}+p_{GG}+p_{GN}+p_{NC}+p_{NG}+p_{NN}$ \\
0 & 0 & 0 & 0 & $1-2p_{A}+p_{AA}$ \\ \hline
1 & 0 & 0 & 0 & $p_{AS}+p_{AC}+p_{AG}+p_{AN}$ \\
1 & 0 & 1 & 1 & $p_{NA}+p_{NS}+p_{NC}+p_{NG}$ \\
1 & 0 & 0 & 1 & $p_{AG}+p_{AN}+p_{SG}+p_{SN}$ \\
1 & 0 & 1 & 0 & $p_{AC}+p_{AG}+p_{AN}+p_{SC}+p_{SG}+p_{SN}+p_{CC}+p_{CG}+p_{CN}$ \\ \hline
0 & 1 & 0 & 0 & $p_{AS}+p_{AC}+p_{AG}+p_{AN}$ \\
0 & 1 & 1 & 1 & $p_{NA}+p_{NS}+p_{NC}+p_{NG}$ \\
0 & 1 & 1 & 0 & $p_{AG}+p_{AN}+p_{SG}+p_{SN}$ \\
0 & 1 & 0 & 1 & $p_{AC}+p_{AG}+p_{AN}+p_{SC}+p_{SG}+p_{SN}+p_{CC}+p_{CG}+p_{CN}$ \\ \hline
\end{tabular}
\end{center}
\end{table}

\begin{table}
\caption{System of equations - simplified}\label{tab:seq2}
\begin{center}
\begin{tabular}{cccccc}
$D_{ig}$ & $D_{jg} $ & $Z_{ig}$ & $Z_{jg}$ & Probabilities & Independent?\\ \hline \hline
1 & 1 & 0 & 0 & $p_{AA}$ & 1\\
1 & 1 & 0 & 1 & $p_{A}+p_{S}-(p_{AG}+p_{AN}+p_{SG}+p_{SN})$ & 2\\
1 & 1 & 1 & 0 & $p_{A}+p_{S}-(p_{AG}+p_{AN}+p_{SG}+p_{SN})$ & - \\
1 & 1 & 1 & 1 & $1-2p_{N}+p_{NN}$  & 3\\ \hline
0 & 0 & 1 & 1 & $p_{NN}$ & 4\\
0 & 0 & 1 & 0 & $p_{G}+p_{N}-(p_{AG}+p_{AN}+p_{SG}+p_{SN})$ & 5 \\
0 & 0 & 0 & 1 & $p_{G}+p_{N}-(p_{AG}+p_{AN}+p_{SG}+p_{SN})$ & -\\
0 & 0 & 0 & 0 & $1-2p_{A}+p_{AA}$ & 6 \\ \hline
1 & 0 & 0 & 0 & $p_{A}-p_{AA}$ & - \\
1 & 0 & 1 & 1 & $p_{N}-p_{NN}$ & - \\
1 & 0 & 0 & 1 & $p_{AG}+p_{AN}+p_{SG}+p_{SN}$ & 7 \\
1 & 0 & 1 & 0 & $p_{C}+(p_{AG}+p_{AN}+p_{SG}+p_{SN})$ & - \\ \hline
0 & 1 & 0 & 0 & $p_{A}-p_{AA}$ & - \\
0 & 1 & 1 & 1 & $p_{N}-p_{NN}$ & - \\
0 & 1 & 1 & 0 & $p_{AG}+p_{AN}+p_{SG}+p_{SN}$ & - \\
0 & 1 & 0 & 1 & $p_{C}+(p_{AG}+p_{AN}+p_{SG}+p_{SN})$ & - \\ \hline
\end{tabular}
\end{center}
\end{table}


\subsection{Proof of Lemma \ref{lemma:direct_ITT}}

Using that:
\begin{align*}
\E[Y_{ig}|Z_{ig}=z,Z_{jg}=z']&=\E[Y_{ig}(0,0)]\\
&+\E[(Y_{ig}(1,0)-Y_{ig}(0,0))D_{ig}(z,z')(1-D_{jg}(z',z))]\\
&+\E[(Y_{ig}(0,1)-Y_{ig}(0,0))(1-D_{ig}(z,z'))D_{jg}(z',z)]\\
&+\E[(Y_{ig}(1,1)-Y_{ig}(0,0))D_{ig}(z,z')D_{jg}(z',z)],
\end{align*}
we have:
\begin{align*}
\E[Y_{ig}|Z_{ig}=1,Z_{jg}&=0]-\E[Y_{ig}|Z_{ig}=0,Z_{jg}=0]=\\
&+\E[(Y_{ig}(1,0)-Y_{ig}(0,0))(D_{ig}(1,0)(1-D_{jg}(0,1))-D_{ig}(0,0)(1-D_{jg}(0,0)))]\\
&+\E[(Y_{ig}(0,1)-Y_{ig}(0,0))((1-D_{ig}(1,0))D_{jg}(0,1)-(1-D_{ig}(0,0))D_{jg}(0,0))]\\
&+\E[(Y_{ig}(1,1)-Y_{ig}(0,0))(D_{ig}(1,0)D_{jg}(0,1)-D_{ig}(0,0)D_{jg}(0,0))].
\end{align*}
Tables \ref{tab:directITT1}, \ref{tab:directITT2} and \ref{tab:directITT3} list the possible values that the terms that depend on the potential treatment statuses can take, which gives the desired result after some algebra. $\square$

\begin{table}
\caption{$D_{ig}(1,0)(1-D_{jg}(0,1))-D_{ig}(0,0)(1-D_{jg}(0,0))$}
\begin{center}
\begin{tabular}{cccc|c|cc}\label{tab:directITT1}
$D_{ig}(1,0)$ & $D_{jg}(0,1)$ & $D_{ig}(0,0)$ & $D_{jg}(0,0)$ & Difference & $\xi_{ig}$  & $\xi_{jg}$\\ \hline \hline
1 & 1 & 1 & 1 & 0  \\
1 & 1 & 1 & 0 & -1 & AT & SC \\
1 & 1 & 0 & 1 & 0  \\
1 & 1 & 0 & 0 & 0  \\
1 & 0 & 1 & 0 & 0  \\
0 & 1 & 0 & 1 & 0  \\
0 & 1 & 0 & 0 & 0  \\
1 & 0 & 0 & 0 & 1 & C,SC & C,GC,NT \\
0 & 0 & 0 & 0 & 0 \\
\end{tabular}
\end{center}
\end{table}

\begin{table}
\caption{$(1-D_{ig}(1,0))D_{jg}(0,1)-(1-D_{ig}(0,0))D_{jg}(0,0)$}
\begin{center}
\begin{tabular}{cccc|c|cc}\label{tab:directITT2}
$D_{ig}(1,0)$ & $D_{jg}(0,1)$ & $D_{ig}(0,0)$ & $D_{jg}(0,0)$ & Difference & $\xi_{ig}$  & $\xi_{jg}$\\ \hline \hline
1 & 1 & 1 & 1 & 0  \\
1 & 1 & 1 & 0 & 0 \\
1 & 1 & 0 & 1 & -1 & C,SC & AT  \\
1 & 1 & 0 & 0 & 0  \\
1 & 0 & 1 & 0 & 0  \\
0 & 1 & 0 & 1 & 0  \\
0 & 1 & 0 & 0 & 1 & GC,NT & SC  \\
1 & 0 & 0 & 0 & 0\\
0 & 0 & 0 & 0 & 0 \\
\end{tabular}
\end{center}
\end{table}

\begin{table}
\caption{$D_{ig}(1,0))D_{jg}(0,1)-D_{ig}(0,0)D_{jg}(0,0)$}
\begin{center}
\begin{tabular}{cccc|c|cc}\label{tab:directITT3}
$D_{ig}(1,0)$ & $D_{jg}(0,1)$ & $D_{ig}(0,0)$ & $D_{jg}(0,0)$ & Difference & $\xi_{ig}$  & $\xi_{jg}$\\ \hline \hline
1 & 1 & 1 & 1 & 0  \\
1 & 1 & 1 & 0 & 1 & AT & SC \\
1 & 1 & 0 & 1 & 1 & C,SC & AT  \\
1 & 1 & 0 & 0 & 1 & C,SC & SC \\
1 & 0 & 1 & 0 & 0  \\
0 & 1 & 0 & 1 & 0  \\
0 & 1 & 0 & 0 & 0 \\
1 & 0 & 0 & 0 & 0\\
0 & 0 & 0 & 0 & 0 \\
\end{tabular}
\end{center}
\end{table}


\subsection{Proof of Proposition \ref{prop:lapo_osnc}}

The result follows using that
\begin{align*}
\E[Y_{ig}|Z_{ig}=z,Z_{jg}=z']&=\E[Y_{ig}(0,0)]\\
&+\E[(Y_{ig}(1,0)-Y_{ig}(0,0))D_{ig}(z,z')(1-D_{jg}(z',z))]\\
&+\E[(Y_{ig}(0,1)-Y_{ig}(0,0))(1-D_{ig}(z,z'))D_{jg}(z',z)]\\
&+\E[(Y_{ig}(1,1)-Y_{ig}(0,0))D_{ig}(z,z')D_{jg}(z',z)],
\end{align*}
combined with the facts that under one-sided noncompliance, $D_{ig}(0,1)=D_{ig}(0,0)=0$, for all $i$, $D_{ig}(1,0)=1$  implies that $i$ is a complier and $D_{ig}(1,1)=0$ implies $i$ is a never-taker (see Table \ref{tab:identif_monot_OSN}). In particular, $\E[Y_{ig}D_{ig}|Z_{ig}=1,Z_{jg}=0]=\E[Y_{ig}(1,0)D_{ig}(1,0)]=\E[Y_{ig}(1,0)|D_{ig}(1,0)=1]\P[D_{ig}(1,0)=1]$ and $\E[Y_{ig}D_{jg}|Z_{ig}=0,Z_{jg}=1]=\E[Y_{ig}(0,1)|D_{jg}(1,0)=1]\P[D_{jg}(1,0)=1]$. On the other hand, $\E[Y_{ig}|Z_{ig}=0,Z_{jg}=0]=\E[Y_{ig}(0,0)]=\E[Y_{ig}(0,0)|C_{ig}]\P[C_{ig}]+\E[Y_{ig}(0,0)|C^c_{ig}]\P[C^c_{ig}]$ and $\E[Y_{ig}(1-D_{ig})|Z_{ig}=1,Z_{jg}=0]=\E[Y_{ig}(0,0)|D_{ig}(1,0)=0]\P[D_{ig}(1,0)=0]=\E[Y_{ig}(0,0)|C^c_{ig}]\P[C^c_{ig}]$. The remaining parts follow analogously. $\square$


\subsection{Proof of Corollary \ref{coro:late_osnc}}

Combine lines 2 and 5 from the display in Proposition \ref{prop:lapo_osnc} and the results in Proposition \ref{prop:comp_distr}, noting that under one-sided noncompliance $\E[D_{ig}|Z_{ig}=0,Z_{jg}=1]=0$. $\square$


\subsection{Proof of Corollary \ref{coro:het_osnc}}

We have that $\E[Y_{ig}(0,0)]=\E[Y_{ig}(0,0)|C_{ig}]\P[C_{ig}]+\E[Y_{ig}(0,0)|C_{ig}^c]\P[C_{ig}^c]$ and thus
\begin{align*}
\E[Y_{ig}(0,0)|C_{ig}^c]&=\frac{\E[Y_{ig}(0,0)]-\E[Y_{ig}(0,0)|C_{ig}]\P[C_{ig}]}{1-\P[C_{ig}]}
\end{align*}
from which
\begin{align*}
\E[Y_{ig}(0,0)|C_{ig}]-\E[Y_{ig}(0,0)|C_{ig}^c]&=\frac{\E[Y_{ig}(0,0)|C_{ig}]-\E[Y_{ig}(0,0)]}{1-\P[C_{ig}]}.
\end{align*}
Using Proposition \ref{prop:lapo_osnc}, we obtain
\begin{multline*}
\E[Y_{ig}(0,0)|C_{ig}]-\E[Y_{ig}(0,0)|C_{ig}^c]=\\
\left\{\E[Y_{ig}|Z_{ig}=0,Z_{jg}=0]-\frac{\E[Y_{ig}(1-D_{ig})|Z_{ig}=1,Z_{jg}=0]}{1-\E[D_{ig}|Z_{ig}=1,Z_{jg}=0]}\right\}\frac{1}{\E[D_{ig}|Z_{ig}=1,Z_{ig}=0]}.
\end{multline*}
Similarly,
\begin{align*}
\E[Y_{ig}(0,0)|C_{jg}]-\E[Y_{ig}(0,0)|C_{jg}^c]&=\frac{\E[Y_{ig}(0,0)|C_{jg}]-\E[Y_{ig}(0,0)]}{1-\P[C_{jg}]}.
\end{align*}
and thus
\begin{multline*}
\E[Y_{ig}(0,0)|C_{jg}]-\E[Y_{ig}(0,0)|C_{jg}^c]=\\
\left\{\E[Y_{ig}|Z_{ig}=0,Z_{jg}=0]-\frac{\E[Y_{ig}(1-D_{jg})|Z_{ig}=0,Z_{jg}=1]}{1-\E[D_{ig}|Z_{ig}=0,Z_{jg}=1]}\right\}\frac{1}{\E[D_{jg}|Z_{ig}=0,Z_{ig}=1]}. \,\square
\end{multline*}


\subsection{Proof of Proposition \ref{prop:spec}}

Under one-sided noncompliance, for any Borel set $\mathcal{Y}$,
\begin{align*}
\P[Y_{ig}\in\mathcal{Y},D_{ig}=0|Z_{ig}=1,Z_{jg}=0]&=\P[Y_{ig}(0,0)\in\mathcal{Y},D_{ig}(1,0)=0|Z_{ig}=1,Z_{jg}=0]\\
&=\P[Y_{ig}(0,0)\in\mathcal{Y},D_{ig}(1,0)=0]\\
&=\P[Y_{ig}(0,0)\in\mathcal{Y},C_{ig}^c]
\end{align*}
and
\begin{align*}
\P[Y_{ig}\in\mathcal{Y}|Z_{ig}=0,Z_{jg}=0]&=\P[Y_{ig}(0,0)\in\mathcal{Y}|Z_{ig}=0,Z_{jg}=0]\\
&=\P[Y_{ig}(0,0)\in\mathcal{Y}]\\
&=\P[Y_{ig}(0,0)\in\mathcal{Y},C_{ig}^c]+\P[Y_{ig}(0,0)\in\mathcal{Y},C_{ig}]
\end{align*}
from which
\begin{align*}
\P[Y_{ig}(0,0)\in\mathcal{Y},C_{ig}]&=\P[Y_{ig}\in\mathcal{Y}|Z_{ig}=0,Z_{jg}=0]-\P[Y_{ig}\in\mathcal{Y},D_{ig}=0|Z_{ig}=1,Z_{jg}=0].
\end{align*}
Since $\P[Y_{ig}(0,0)\in\mathcal{Y},C_{ig}]\ge 0$, a testable implication of this model is:
\[\P[Y_{ig}\in\mathcal{Y}|Z_{ig}=0,Z_{jg}=0]-\P[Y_{ig}\in\mathcal{Y},D_{ig}=0|Z_{ig}=1,Z_{jg}=0]\ge 0.\]
By the same reasoning, 
\begin{align*}
\P[Y_{ig}(0,0)\in\mathcal{Y},C_{jg}]&=\P[Y_{ig}\in\mathcal{Y}|Z_{ig}=0,Z_{jg}=0]-\P[Y_{ig}\in\mathcal{Y},D_{jg}=0|Z_{ig}=0,Z_{jg}=1]
\end{align*}
and the testable implication is;
\[\P[Y_{ig}\in\mathcal{Y}|Z_{ig}=0,Z_{jg}=0]-\P[Y_{ig}\in\mathcal{Y},D_{jg}=0|Z_{ig}=0,Z_{jg}=1]\ge 0\]
as required. $\square$


\subsection{Proof of Theorem \ref{thm:2SLS}}

First, consider the 2SLS regression of $Y_{ig}$ on $1-D_{ig}$ and $D_{ig}$ (without an intercept) using $Z_{ig}$ and $1-Z_{ig}$ as instruments, on the subsample of units with $Z_{jg}=0$. The 2SLS estimator is given by:
\[\boldsymbol{\hat{\alpha}}=(\mathbf{\tilde{Z}}'\mathbf{\tilde{D}})^{-1}\mathbf{\tilde{Z}}\mathbf{Y}=\left(\sum_{g,i}\begin{bmatrix}
1-Z_{ig} \\ 
Z_{ig}
\end{bmatrix}\begin{bmatrix}
1-D_{ig} & D_{ig}
\end{bmatrix} (1-Z_{jg})\right)^{-1}\sum_{g,i}\begin{bmatrix}
1-Z_{ig} \\ 
Z_{ig}
\end{bmatrix} Y_{ig}(1-Z_{jg})=\begin{bmatrix}
\bar{Y}_{00}\\\
\frac{\bar{Y}_{10}-\bar{Y}_{00}}{\bar{D}_{10}}+\bar{Y}_{00}
\end{bmatrix}.\]
The cluster-robust variance estimator is:
\[\hat{V}_{\mathsf{cr}}(\boldsymbol{\hat{\alpha}})=\frac{1}{4G^2}\left(\frac{1}{2G}\mathbf{\tilde{Z}}'\mathbf{\tilde{D}}\right)^{-1}\sum_g \mathbf{\tilde{z}}_g'\mathbf{\hat{u}}_g\mathbf{\hat{u}}_g'\mathbf{\tilde{z}}_g\left(\frac{1}{2G}\mathbf{\tilde{D}}'\mathbf{\tilde{Z}}\right)^{-1}\]
where $\hat{u}_{ig}=(Y_{ig}-\hat{\alpha}_0(1-D_{ig})-\hat{\alpha}_1D_{ig})(1-Z_{jg})$ and $\mathbf{\hat{u}}_g'=[\hat{u}_{1g}\,  \hat{u}_{2g}]$.
Next,
\begin{align*}
\mathbf{\tilde{z}}_g'\mathbf{\hat{u}}_g\mathbf{\hat{u}}_g'\mathbf{\tilde{z}}_g&=\begin{bmatrix}
(1-Z_{1g})(1-Z_{2g})(\hat{u}_{1g}^2+\hat{u}_{2g}^2+2\hat{u}_{1g}\hat{u}_{2g}) & 0 \\
0 & Z_{1g}(1-Z_{2g})\hat{u}_{1g}^2+Z_{2g}(1-Z_{1g})\hat{u}_{2g}^2
\end{bmatrix}
\end{align*}
Then,
\[\hat{V}_{\mathsf{cr},11}(\boldsymbol{\hat{\alpha}})=\frac{1}{4G^2\hat{p}_{00}^2}\sum_{g,i}(1-Z_{ig})(1-Z_{jg})(\hat{u}_{ig}^2+\hat{u}_{ig}\hat{u}_{jg})\]
and
\begin{align*}
\hat{V}_{\mathsf{cr},22}(\boldsymbol{\hat{\alpha}})&=\frac{1}{4G^2\bar{D}_{10}^2\hat{p}_{10}^2}\sum_{g,i}Z_{ig}(1-Z_{jg})\hat{u}_{ig}^2+\frac{(1-\bar{D}_{10})^2}{4G^2\bar{D}_{10}^2\hat{p}_{00}^2}\sum_{g,i}(1-Z_{ig})(1-Z_{jg})(\hat{u}_{ig}^2+\hat{u}_{ig}\hat{u}_{jg}).
\end{align*}
Now, for any invertible transformation $\mathbf{\tilde{A}}$, consider the transformed variables $\mathbf{\tilde{Z}}\mathbf{\tilde{A}}$ and $\mathbf{\tilde{D}}\mathbf{\tilde{A}}$. Let: $\boldsymbol{\hat{\delta}}(\mathbf{\tilde{A}})=((\mathbf{\tilde{Z}}\mathbf{\tilde{A}})'(\mathbf{\tilde{D}}\mathbf{\tilde{A}}))^{-1}(\mathbf{\tilde{Z}}\mathbf{\tilde{A}})'\mathbf{Y}$. The 2SLS estimator using this transformed variables is:
\begin{align*}
\boldsymbol{\hat{\delta}}(\mathbf{\tilde{A}})&=((\mathbf{\tilde{Z}}\mathbf{\tilde{A}})'(\mathbf{\tilde{D}}\mathbf{\tilde{A}}))^{-1}(\mathbf{\tilde{Z}}\mathbf{\tilde{A}})'\mathbf{Y}\\
&=(\mathbf{\tilde{A}}'\mathbf{\tilde{Z}}'\mathbf{\tilde{D}}\mathbf{\tilde{A}})^{-1}\mathbf{\tilde{A}}'\mathbf{\tilde{Z}}'\mathbf{Y}\\
&=\mathbf{\tilde{A}}^{-1}(\mathbf{\tilde{Z}}'\mathbf{\tilde{D}})^{-1}(\mathbf{\tilde{A}}')^{-1}\mathbf{\tilde{A}}'\mathbf{\tilde{Z}}'\mathbf{Y} \\
&=\mathbf{\tilde{A}}^{-1}(\mathbf{\tilde{Z}}'\mathbf{\tilde{D}})^{-1}\mathbf{\tilde{Z}}'\mathbf{Y}\\
&=\mathbf{\tilde{A}}^{-1}\boldsymbol{\hat{\alpha}}.
\end{align*}
and 
\[\hat{V}_{\mathsf{cr}}(\boldsymbol{\hat{\delta}}(\mathbf{\tilde{A}}))=\mathbf{\tilde{A}}^{-1}\hat{V}_{\mathsf{cr}}(\boldsymbol{\hat{\alpha}})(\mathbf{\tilde{A}}^{-1})'\]
Now, setting:
\[\mathbf{\tilde{A}}=\begin{bmatrix}
1 & 0 \\
1 & 1 
\end{bmatrix}\]
gives the 2SLS estimator from the regression of $Y_{ig}$ on $D_{ig}$, and a constant using $Z_{ig}$ as an instrument, conditional on $Z_{jg}=0$, which is the parameterization considered in the paper. It follows that the conditional Wald estimator of the direct effect on compliers is:
\[\hat{\delta}_1=\frac{\bar{Y}_{10}-\bar{Y}_{00}}{\bar{D}_{10}}\]
and 
\[\hat{V}_{\mathsf{cr}}(\hat{\delta}_1)=\hat{V}_{\mathsf{cr},11}(\boldsymbol{\hat{\alpha}})+\hat{V}_{\mathsf{cr},22}(\boldsymbol{\hat{\alpha}}).\]

The case for the spillover effect estimator based on the regression of $Y_{ig}$ on $D_{jg}$ using $Z_{jg}$ as an instrument on the subsample of units with $Z_{ig}=0$ follows analogously.

Next, consider the 2SLS regression of $Y_{ig}$ on $(1-D_{ig})(1-D_{jg})$, $D_{ig}(1-D_{jg})$, $(1-D_{ig})D_{jg}$ and $D_{ig}D_{jg}$ using $(1-Z_{ig})(1-Z_{jg})$, $Z_{ig}(1-Z_{jg})$, $(1-Z_{ig})Z_{jg}$ and $Z_{ig}Z_{jg}$ as instruments. The 2SLS estimator is:
\begin{align*}
\boldsymbol{\hat{\theta}}=\left(\frac{1}{2G}\mathbf{Z}'\mathbf{D}\right)^{-1}\frac{1}{2G}\mathbf{Z}'\mathbf{Y}.
\end{align*}
Now, 
\begin{align*}
\frac{1}{2G}\mathbf{Z}'\mathbf{D}&=\frac{1}{2G}\sum_{g,i}\begin{bmatrix}
(1-Z_{ig})(1-Z_{jg}) \\
Z_{ig}(1-Z_{jg}) \\
(1-Z_{ig})Z_{jg}\\
Z_{ig}Z_{jg}
\end{bmatrix}  \begin{bmatrix}
(1-D_{ig})(1-D_{jg}) & D_{ig}(1-D_{jg}) & (1-D_{ig})D_{jg} & D_{ig}D_{jg}
\end{bmatrix}\\
&=\begin{bmatrix}
\hat{p}_{00} & 0 & 0 & 0 \\
(1-\bar{D}_{10})\hat{p}_{10} & \bar{D}_{10}\hat{p}_{10} & 0 & 0 \\
(1-\bar{D}_{10})\hat{p}_{10} & 0 & \bar{D}_{10}\hat{p}_{10} & 0 \\
\bar{D}_{11}^{00}\hat{p}_{11} & \bar{D}_{11}^{10}\hat{p}_{11} & \bar{D}_{11}^{01}\hat{p}_{11} & \bar{D}_{11}^{11}\hat{p}_{11}
\end{bmatrix}
\end{align*}
where 
\begin{align*}
\hat{p}_{zz'}&=\frac{1}{2G}\sum_{g,i}\1(Z_{ig}=z)\1(Z_{jg}=z') \\
\bar{D}_{z,z'}&=\frac{\sum_{g,i}D_{ig}\1(Z_{ig}=z)\1(Z_{jg}=z')}{\sum_{g,i}\1(Z_{ig}=z)\1(Z_{jg}=z')} \\
\bar{D}_{z,z'}^{d,d'}&=\frac{\sum_{g,i}\1(D_{ig}=d)\1(D_{jg}=d')\1(Z_{ig}=z)\1(Z_{jg}=z')}{\sum_{g,i}\1(Z_{ig}=z)\1(Z_{jg}=z')}.
\end{align*}
The inverse can be found by direct calculation as the matrix $\mathbf{\hat{Q}}$ such that $\frac{1}{2G}\mathbf{Z}'\mathbf{D}\times \mathbf{\hat{Q}}=\mathbf{I}$. This gives:
\begin{align*}
\mathbf{\hat{Q}}=\left(\frac{1}{2G}\mathbf{Z}'\mathbf{D}\right)^{-1}&=\begin{bmatrix}
\frac{1}{\hat{p}_{00}} & 0 & 0 & 0 \\
-\frac{(1-\bar{D}_{10})}{\bar{D}_{10}\hat{p}_{00}} & \frac{1}{\bar{D}_{10}\hat{p}_{10}} & 0 & 0 \\
-\frac{(1-\bar{D}_{10})}{\bar{D}_{10}\hat{p}_{00}} & 0 & \frac{1}{\bar{D}_{10}\hat{p}_{10}} & 0 \\
\frac{2\bar{D}_{11}^{10}}{\hat{p}_{00}\bar{D}_{11}^{11}}\left(\frac{1-\bar{D}_{10}}{\bar{D}_{10}}\right)-\frac{\bar{D}_{11}^{00}}{\hat{p}_{00}\bar{D}_{11}^{11}}& -\frac{\bar{D}_{11}^{10}}{\bar{D}_{10}\bar{D}_{11}^{11}\hat{p}_{10}} & -\frac{\bar{D}_{11}^{10}}{\bar{D}_{10}\bar{D}_{11}^{11}\hat{p}_{10}} & \frac{1}{\bar{D}_{11}^{11}\hat{p}_{11}}
\end{bmatrix}.
\end{align*}
On the other hand,
\begin{align*}
\frac{1}{2G}\mathbf{Z}'\mathbf{Y}&=\begin{bmatrix}
\bar{Y}_{00}\hat{p}_{00}\\
\bar{Y}_{10}\hat{p}_{10}\\
\bar{Y}_{01}\hat{p}_{10}\\
\bar{Y}_{11}\hat{p}_{11}
\end{bmatrix}
\end{align*}
where
\[\bar{Y}_{zz'}=\frac{\sum_{g,i}Y_{ig}\1(Z_{ig}=z)\1(Z_{jg}=z')}{\sum_{g,i}\1(Z_{ig}=z)\1(Z_{jg}=z')}.\]
Thus:
\begin{align*}
\boldsymbol{\hat{\theta}}&=\begin{bmatrix}
\bar{Y}_{00}\\
\frac{\bar{Y}_{10}-\bar{Y}_{00}}{\bar{D}_{10}}+\bar{Y}_{00} \\
\frac{\bar{Y}_{01}-\bar{Y}_{00}}{\bar{D}_{10}}+\bar{Y}_{00} \\
\frac{\bar{Y}_{11}-\bar{Y}_{00}}{\bar{D}_{11}^{11}}-\frac{\bar{D}_{11}^{10}}{\bar{D}_{11}^{11}}\left(\frac{\bar{Y}_{10}+\bar{Y}_{01}-2\bar{Y}_{00}}{\bar{D}_{10}}\right)+\bar{Y}_{00}
\end{bmatrix}.
\end{align*}
Now, for any invertible transformation $\mathbf{A}$, consider the transformed variables $\mathbf{Z}\mathbf{A}$ and $\mathbf{D}\mathbf{A}$. Let: $\boldsymbol{\hat{\beta}}(\mathbf{A})=((\mathbf{Z}\mathbf{A})'(\mathbf{D}\mathbf{A}))^{-1}(\mathbf{Z}\mathbf{A})'\mathbf{Y}$. The 2SLS estimator using this transformed variables is:
\begin{align*}
\boldsymbol{\hat{\beta}}(\mathbf{A})&=((\mathbf{Z}\mathbf{A})'(\mathbf{D}\mathbf{A}))^{-1}(\mathbf{Z}\mathbf{A})'\mathbf{Y}\\
&=(\mathbf{A}'\mathbf{Z}'\mathbf{D}\mathbf{A})^{-1}\mathbf{A}'\mathbf{Z}'\mathbf{Y}\\
&=\mathbf{A}^{-1}(\mathbf{Z}'\mathbf{D})^{-1}(\mathbf{A}')^{-1}\mathbf{A}'\mathbf{Z}'\mathbf{Y} \\
&=\mathbf{A}^{-1}(\mathbf{Z}'\mathbf{D})^{-1}\mathbf{Z}'\mathbf{Y}\\
&=\mathbf{A}^{-1}\boldsymbol{\hat{\theta}}.
\end{align*}
and 
\[\hat{V}_{\mathsf{cr}}(\boldsymbol{\hat{\beta}}(\mathbf{A}))=\mathbf{A}^{-1}\hat{V}_{\mathsf{cr}}(\boldsymbol{\hat{\theta}})(\mathbf{A}^{-1})'\]
Now, setting:
\[\mathbf{A}=\begin{bmatrix}
1 & 0 & 0 & 0 \\
1 & 1 & 0 & 0 \\
1 & 0 & 1 & 0 \\
1 & 1 & 1 & 1
\end{bmatrix}\]
gives the 2SLS estimator from the regression of $Y_{ig}$ on $D_{ig}$, $D_{jg}$, $D_{ig}D_{jg}$ and a constant using $Z_{ig}$, $Z_{jg}$, $Z_{ig}Z_{jg}$ as instruments, which is the parameterization considered in the paper. It follows that the 2SLS estimator, $\boldsymbol{\hat{\beta}}$, is:
\begin{align*}
\boldsymbol{\hat{\beta}}&=\begin{bmatrix} 
\hat{\beta}_0 \\
\hat{\beta}_1 \\
\hat{\beta}_2 \\
\hat{\beta}_3
\end{bmatrix}=\begin{bmatrix}
1 & 0 & 0 & 0 \\
-1 & 1 & 0 & 0 \\
-1 & 0 & 1 & 0 \\
1 & -1 & -1 & 1
\end{bmatrix}\boldsymbol{\hat{\theta}}\\
&=\begin{bmatrix}
\bar{Y}_{00}\\
\frac{\bar{Y}_{10}-\bar{Y}_{00}}{\bar{D}_{10}} \\
\frac{\bar{Y}_{01}-\bar{Y}_{00}}{\bar{D}_{10}}\\
\frac{\bar{Y}_{11}-\bar{Y}_{00}}{\bar{D}_{11}^{11}}-\left(\frac{\bar{D}_{11}^{11}+\bar{D}_{11}^{10}}{\bar{D}_{11}^{11}}\right)\left(\frac{\bar{Y}_{10}-\bar{Y}_{00}}{\bar{D}_{10}}+\frac{\bar{Y}_{01}-\bar{Y}_{00}}{\bar{D}_{10}}\right)
\end{bmatrix}
\end{align*}
which implies that the treatment effect estimators $\hat{\beta}_1$ and $\hat{\beta}_2$ are identical to the conditional Wald ratio estimators. 

The cluster-robust variance estimator for $\boldsymbol{\hat{\theta}}$ is:
\[\hat{V}_\mathsf{cr}(\boldsymbol{\hat{\theta}})=\frac{1}{4G^2}\left(\frac{1}{2G}\mathbf{Z}'\mathbf{D}\right)^{-1} \sum_g \mathbf{z}_g'\boldsymbol{\hat{\varepsilon}}_g\boldsymbol{\hat{\varepsilon}}_g'\mathbf{z}_g\left(\frac{1}{2G}\mathbf{D}'\mathbf{Z}\right)^{-1}\]
where $\hat{\varepsilon}_{ig}=Y_{ig}-\mathbf{D}_{ig}'\boldsymbol{\hat\theta}$.
We have that:
\begin{align*}
\mathbf{z}_g'\boldsymbol{\hat{\varepsilon}}_g\boldsymbol{\hat{\varepsilon}}_g'&=\begin{bmatrix}
(1-Z_{1g})(1-Z_{2g})(\hat{\varepsilon}_{1g}^2+\hat{\varepsilon}_{1g}\hat{\varepsilon}_{2g}) & (1-Z_{2g})(1-Z_{1g})(\hat{\varepsilon}_{2g}^2+\hat{\varepsilon}_{1g}\hat{\varepsilon}_{2g}) \\
Z_{1g}(1-Z_{2g})\hat{\varepsilon}_{1g}^2+Z_{2g}(1-Z_{1g})\hat{\varepsilon}_{1g}\hat{\varepsilon}_{2g} & Z_{2g}(1-Z_{1g})\hat{\varepsilon}_{2g}^2+Z_{1g}(1-Z_{2g})\hat{\varepsilon}_{1g}\hat{\varepsilon}_{2g} \\
(1-Z_{1g})Z_{2g}\hat{\varepsilon}_{1g}^2 + (1-Z_{2g})Z_{1g}\hat{\varepsilon}_{1g}\hat{\varepsilon}_{2g} & (1-Z_{2g})Z_{1g}\hat{\varepsilon}_{2g}^2 + (1-Z_{1g})Z_{2g}\hat{\varepsilon}_{1g}\hat{\varepsilon}_{2g} \\
Z_{1g}Z_{2g}(\hat{\varepsilon}_{1g}^2+\hat{\varepsilon}_{1g}\hat{\varepsilon}_{2g}) & Z_{1g}Z_{2g}(\hat{\varepsilon}_{1g}^2+\hat{\varepsilon}_{1g}\hat{\varepsilon}_{2g})
\end{bmatrix}
\end{align*}
and
\begin{align*}
\boldsymbol{\hat{\Omega}}&=\sum_g\mathbf{z}_g'\boldsymbol{\hat{\varepsilon}}_g\boldsymbol{\hat{\varepsilon}}_g'\mathbf{z}_g\\
&=\begin{bmatrix}
\omega_{11} & 0 & 0 & 0 \\
0 & \omega_{22} & \omega_{23} & 0 \\
0 & \omega_{23} & \omega_{33} & 0 \\
0 & 0 & 0 & \omega_{44}
\end{bmatrix}
\end{align*}
where
\begin{align*}
\omega_{11}&=\sum_{g,i}(1-Z_{ig})(1-Z_{jg})(\hat{\varepsilon}_{ig}^2+\hat{\varepsilon}_{ig}\hat{\varepsilon}_{jg}) \\
\omega_{22}&=\sum_{g,i}Z_{ig}(1-Z_{jg})\hat{\varepsilon}_{ig}^2\\
\omega_{23}&=\sum_{g,i}Z_{ig}(1-Z_{jg})\hat{\varepsilon}_{ig}\hat{\varepsilon}_{jg} \\
\omega_{33}&=\sum_{g,i}(1-Z_{ig})Z_{jg}\hat{\varepsilon}_{ig}^2\\
\omega_{44}&=\sum_{g,i}Z_{ig}Z_{jg}(\hat{\varepsilon}_{ig}^2+\hat{\varepsilon}_{ig}\hat{\varepsilon}_{jg}).
\end{align*}
Let
\[\mathbf{\hat{Q}}=\left(\frac{1}{2G}\mathbf{Z}'\mathbf{D}\right)^{-1}=\begin{bmatrix}
q_{11} & 0 & 0 & 0 \\
q_{21} & q_{22} & 0 & 0 \\
q_{21} & 0 & q_{22} & 0 \\
q_{41} & q_{42} & q_{43} & q_{44}
\end{bmatrix}\]
where the exact values of the matrix were given above. The cluster-robust variance estimator for $\boldsymbol{\hat{\theta}}$ can be rewritten as:
\[\hat{V}_\mathsf{cr}(\boldsymbol{\hat{\theta}})=\frac{1}{4G^2}\mathbf{\hat{Q}}\boldsymbol{\hat{\Omega}}\mathbf{\hat{Q}}'.\]
Then,
\begin{align*}
\hat{V}_{\mathsf{cr},11}(\boldsymbol{\hat{\theta}})&=\frac{1}{4G^2\hat{p}_{00}^2}\sum_{g,i}(1-Z_{ig})(1-Z_{jg})(\hat{\varepsilon}_{ig}^2+\hat{\varepsilon}_{ig}\hat{\varepsilon}_{jg})\\
\hat{V}_{\mathsf{cr},22}(\boldsymbol{\hat{\theta}})&=\frac{1}{4G^2\bar{D}_{10}^2\hat{p}_{10}^2}\sum_{g,i}Z_{ig}(1-Z_{jg})\hat{\varepsilon}_{ig}^2+\frac{(1-\bar{D}_{10})^2}{4G^2\bar{D}_{10}^2\hat{p}_{00}^2}\sum_{g,i}(1-Z_{ig})(1-Z_{jg})(\hat{\varepsilon}_{ig}^2+\hat{\varepsilon}_{ig}\hat{\varepsilon}_{jg}).
\end{align*}
But notice that, if the residuals are the same, we have that $\hat{V}_{\mathsf{cr},11}(\boldsymbol{\hat{\theta}})=\hat{V}_{\mathsf{cr},11}(\boldsymbol{\hat{\alpha}})$ and $\hat{V}_{\mathsf{cr},22}(\boldsymbol{\hat{\theta}})=\hat{V}_{\mathsf{cr},22}(\boldsymbol{\hat{\alpha}})$ which in turns implies that the cluster-robust variance estimators for $\hat{\beta}_1$ and $\hat{\delta}_1$ are equal.

To see that the residuals are indeed equal, note that:
\begin{align*}
(1-Z_{jg})\hat{u}_{ig}^2&=(1-Z_{jg})(Y_{ig}-\hat{\alpha}_0(1-D_{ig})-\hat{\alpha}_1D_{ig})\\
&=(1-Z_{jg})(Y_{ig}-\hat{\theta}_0(1-D_{ig})-\hat{\theta}_1D_{ig})\\
&=(1-Z_{jg})\hat{\varepsilon}_{ig}^2
\end{align*}
which implies that $\hat{\delta}_0=\hat{\beta}_0$, $\hat{V}_{\mathsf{cr}}(\hat{\delta}_0)=\hat{V}_{\mathsf{cr}}(\hat{\beta}_0)$, $\hat{\delta}_1=\hat{\beta}_1$ and $\hat{V}_{\mathsf{cr}}(\hat{\delta}_1)=\hat{V}_{\mathsf{cr}}(\hat{\beta}_1)$. The results for $\hat{\delta}_2$ and $\hat{\beta}_2$ follow by the same argument. $\square$


\subsection{Proof of Lemma \ref{lemma:asynorm}}

This result is well-known and follows from standard 2SLS properties and using Theorem \ref{thm:2SLS} as $G\to\infty$. To find the exact formula for $\beta_3$, note that
\begin{align*}
\E[Y_{ig}|Z_{ig}=z,Z_{jg}=z']=\beta_0&+\beta_1\E[D_{ig}|Z_{ig}=z,Z_{jg}=z']\\
&+\beta_2\E[D_{jg}|Z_{ig}=z,Z_{jg}=z']\\
&+\beta_3\E[D_{ig}D_{jg}|Z_{ig}=z,Z_{jg}=z']
\end{align*}

Under one-sided noncompliance, $\E[D_{ig}|Z_{ig}=0,Z_{jg}=z']=\E[D_{jg}|Z_{ig}=z,Z_{jg}=0]=0$ and thus:
\begin{align*}
\E[Y_{ig}|Z_{ig}=0,Z_{jg}=0]&=\beta_0\\
\E[Y_{ig}|Z_{ig}=1,Z_{jg}=0]&=\beta_0+\beta_1\E[D_{ig}|Z_{ig}=1,Z_{jg}=0]\\
\E[Y_{ig}|Z_{ig}=0,Z_{jg}=1]&=\beta_0+\beta_2\E[D_{ig}|Z_{ig}=0,Z_{jg}=1]
\end{align*}
from which:
\begin{align*}
\beta_0&=\E[Y_{ig}|Z_{ig}=0,Z_{jg}=0]=\E[Y_{ig}(0,0)]\\
\beta_1&=\frac{\E[Y_{ig}|Z_{ig}=1,Z_{jg}=0]-\E[Y_{ig}|Z_{ig}=0,Z_{jg}=0]}{\E[D_{ig}|Z_{ig}=1,Z_{jg}=0]}=\E[Y_{ig}(1,0)-Y_{ig}(0,0)|C_{ig}]\\
\beta_2&=\frac{\E[Y_{ig}|Z_{ig}=0,Z_{jg}=1]-\E[Y_{ig}|Z_{ig}=0,Z_{jg}=0]}{\E[D_{jg}|Z_{ig}=0,Z_{jg}=1]}=\E[Y_{ig}(0,1)-Y_{ig}(0,0)|C_{jg}]\\
\beta_3&=\frac{\E[Y_{ig}|Z_{ig}=1,Z_{jg}=1]-\beta_0-\beta_1\E[D_{ig}(1,1)]-\beta_2\E[D_{jg}(1,1)]}{\E[D_{jg}(1,1)D_{jg}(1,1)]}
\end{align*}
as long as $\E[D_{jg}(1,1)D_{jg}(1,1)]>0$ (otherwise, $\beta_3$ is not identified). Finally, note that
\begin{align*}
\E[Y_{ig}|Z_{ig}=1,Z_{jg}=1]&=\E[Y_{ig}(0,0)]\\
&+\E[Y_{ig}(1,0)-Y_{ig}(0,0)|D_{ig}(1,1)=1]\E[D_{ig}(1,1)]\\
&+\E[Y_{ig}(0,1)-Y_{ig}(0,0)|D_{jg}(1,1)=1]\E[D_{jg}(1,1)]\\
&+\E[Y_{ig}(1,1)-Y_{ig}(1,0)-Y_{ig}(0,1)+Y_{ig}(0,0)|D_{ig}(1,1)=1,D_{jg}(1,1)=1]\\
&\quad \times \E[D_{ig}(1,1)D_{jg}(1,1)]
\end{align*}
and use the fact that $D_{ig}(1,1)=1$ if $i$ is a complier or a group complier to get that:
\begin{align*}
\beta_3&=(\E[Y_{ig}(1,0)-Y_{ig}(0,0)|GC_{ig}]-\E[Y_{ig}(1,0)-Y_{ig}(0,0)|C_{ig}])\frac{\P[GC_{ig}]}{\E[D_{ig}(1,1)D_{jg}(1,1)]}\\
&+(\E[Y_{ig}(0,1)-Y_{ig}(0,0)|GC_{jg}]-\E[Y_{ig}(0,1)-Y_{ig}(0,0)|C_{jg}])\frac{\P[GC_{jg}]}{\E[D_{ig}(1,1)D_{jg}(1,1)]}\\
&+\E[Y_{ig}(1,1)-Y_{ig}(1,0)-(Y_{ig}(0,1)-Y_{ig}(0,0))|D_{ig}(1,1)=1,D_{jg}(1,1)=1].
\end{align*}
which gives the desired result. $\square$


\subsection{Proof of Proposition \ref{prop:cia}}

First, 
\begin{align*}
\E[D_{ig}|Z_{ig}=1,Z_{jg}=0,X_g]&=\E[D_{ig}(1,0)|Z_{ig}=1,Z_{jg}=0,X_g]\\
&=\E[D_{ig}(1,0)|X_g]=\P[C_{ig}|X_g]
\end{align*}
and
\begin{align*}
\E[D_{ig}|Z_{ig}=1,Z_{jg}=1,X_g]&=\E[D_{ig}(1,1)|Z_{ig}=1,Z_{jg}=1,X_g]\\
&=\E[D_{ig}(1,1)|X_g]=\P[C_{ig}|X_g]+\P[GC_{ig}|X_g].
\end{align*}
For the second part, we have that for the first term,
\begin{align*}
\E\left[g(Y_{ig},X_g)\frac{(1-Z_{ig})(1-Z_{jg})}{p_{00}(X_g)}\right]&= \E_{X_g}\left\{\E\left[\left.g(Y_{ig},X_g)\frac{(1-Z_{ig})(1-Z_{jg})}{p_{00}(X_g)}\right\vert X_g\right]\right\} \\
&=\E_{X_g}\left\{\E\left[\left.g(Y_{ig},X_g)\right\vert Z_{ig}=0,Z_{jg}=0,X_g\right]\right\} \\
&=\E_{X_g}\left\{\E\left[\left.g(Y_{ig}(0,0),X_g)\right\vert Z_{ig}=0,Z_{jg}=0,X_g\right]\right\} \\
&=\E_{X_g}\left\{\E\left[\left.g(Y_{ig}(0,0),X_g)\right\vert X_g\right]\right\} \\
&=\E[g(Y_{ig}(0,0),X_g)].
\end{align*}
For the second term,
\begin{align*}
\E\left[g(Y_{ig},X_g)D_{ig}\frac{Z_{ig}(1-Z_{ig})}{p_{10}(X_g)}\right]&=\E_{X_g}\left\{\E\left[\left.g(Y_{ig},X_g)D_{ig}\frac{Z_{ig}(1-Z_{ig})}{p_{10}(X_g)}\right\vert X_g\right]\right\}\\
&=\E_{X_g}\left\{\E\left[\left.g(Y_{ig},X_g)D_{ig}\right\vert Z_{ig}=1,Z_{jg}=0,X_g\right]\right\}\\
&=\E_{X_g}\left\{\E\left[\left.g(Y_{ig}(1,0),X_g)D_{ig}(1,0)\right\vert Z_{ig}=1,Z_{jg}=0,X_g\right]\right\}\\
&=\E_{X_g}\left\{\E\left[\left.g(Y_{ig}(1,0),X_g)D_{ig}(1,0)\right\vert X_g\right]\right\}\\
&=\E\left[g(Y_{ig}(1,0),X_g)D_{ig}(1,0)\right]\\
&=\E\left[g(Y_{ig}(1,0),X_g)|C_{ig}\right]\P[C_{ig}].
\end{align*}
For the third term,
\begin{align*}
\E\left[g(Y_{ig},X_g)D_{jg}\frac{(1-Z_{ig})Z_{ig}}{p_{01}(X_g)}\right]&=\E_{X_g}\left\{\E\left[\left.g(Y_{ig},X_g)D_{jg}\frac{(1-Z_{ig})Z_{ig}}{p_{01}(X_g)}\right\vert X_g\right]\right\}\\
&=\E_{X_g}\left\{\E\left[\left.g(Y_{ig},X_g)D_{jg}\right\vert Z_{ig}=0,Z_{jg}=1,X_g\right]\right\} \\
&=\E_{X_g}\left\{\E\left[\left.g(Y_{ig}(0,1),X_g)D_{jg}(0,1)\right\vert Z_{ig}=0,Z_{jg}=1,X_g\right]\right\} \\
&=\E_{X_g}\left\{\E\left[g(Y_{ig}(0,1),X_g)D_{jg}(0,1)\right]\right\} \\
&=\E\left[g(Y_{ig}(0,1),X_g)D_{jg}(0,1)\right]\\
&=\E[g(Y_{ig}(0,1),X_g)|C_{jg}]\P[C_{jg}].
\end{align*}
For the fourth term,
\begin{align*}
\E\left[g(Y_{ig},X_g)(1-D_{ig})\frac{Z_{ig}(1-Z_{jg})}{p_{10}(X_g)}\right]&=\E_{X_g}\left\{\E\left[\left.g(Y_{ig},X_g)(1-D_{ig})\frac{Z_{ig}(1-Z_{jg})}{p_{10}(X_g)}\right\vert X_g\right]\right\}\\
&=\E_{X_g}\left\{\E\left[\left.g(Y_{ig},X_g)(1-D_{ig})\right\vert Z_{ig}=1,Z_{jg}=0,X_g\right]\right\} \\
&=\E_{X_g}\left\{\E\left[\left.g(Y_{ig}(0,0),X_g)(1-D_{ig}(1,0))\right\vert Z_{ig}=1,Z_{jg}=0,X_g\right]\right\} \\
&=\E_{X_g}\left\{\E\left[\left.g(Y_{ig}(0,0),X_g)(1-D_{ig}(1,0))\right\vert X_g\right]\right\} \\
&=\E[g(Y_{ig}(0,0),X_g)(1-D_{ig}(1,0))]\\
&=\E[g(Y_{ig}(0,0),X_g)|C_{ig}^c]\P[C_{ig}^c]
\end{align*}
and the result follows from $\E[g(Y_{ig}(0,0),X_g)]=\E[g(Y_{ig}(0,0),X_g)|C_{ig}]\P[C_{ig}]+\E[g(Y_{ig}(0,0),X_g)|C_{ig}^c]\P[C_{ig}^c]$.
Similarly for the fifth term,
\begin{align*}
\E\left[g(Y_{ig},X_g)(1-D_{jg})\frac{(1-Z_{ig})Z_{jg}}{p_{01}(X_g)}\right] &=\E_{X_g}\left\{\E\left[\left.g(Y_{ig},X_g)(1-D_{jg})\frac{(1-Z_{ig})Z_{jg}}{p_{01}(X_g)}\right\vert X_g\right]\right\}\\
&=\E_{X_g}\left\{\E\left[\left.g(Y_{ig},X_g)(1-D_{jg})\right\vert Z_{ig}=0,Z_{jg}=1,X_g\right]\right\}\\
&=\E_{X_g}\left\{\E\left[\left.g(Y_{ig}(0,0),X_g)(1-D_{jg}(1,0))\right\vert Z_{ig}=0,Z_{jg}=1,X_g\right]\right\}\\
&=\E_{X_g}\left\{\E\left[\left.g(Y_{ig}(0,0),X_g)(1-D_{jg}(1,0))\right\vert X_g\right]\right\}\\
&=\E[g(Y_{ig}(0,0),X_g)(1-D_{jg}(1,0))]\\
&=\E[g(Y_{ig}(0,0),X_g)|C_{jg}^c]\P[C_{jg}^c]
\end{align*}
and it can be seen that all these equalities also hold conditional on $X_g$. $\square$


\subsection{Proof of Proposition \ref{prop:distr_comp_g}}

By independence, $\E[D_{ig}|Z_{ig}=z,W_{ig}=w]=\E[D_{ig}(z,w)]$. Then, under monotonicity, $\E[D_{ig}|Z_{ig}=0,W_{ig}=0]=\E[D_{ig}(0,0)]=\P[D_{ig}(0,0)=1]=\P[AT_{ig}]$. Next, $\E[D_{ig}|Z_{ig}=0,W_{ig}=w^*]-\E[D_{ig}|Z_{ig}=0,W_{ig}=w^*-1]=\E[D_{ig}(0,w^*)]-\E[D_{ig}(0,w^*-1)]=\P[D_{ig}(0,w^*)>D_{ig}(0,w^*-1)]=\P[SC_{ig}(w^*)]$. Similarly, $\E[D_{ig}|Z_{ig}=1,W_{ig}=0]-\E[D_{ig}|Z_{ig}=0,W_{ig}=n_g]=\P[D_{ig}(1,0)>D_{ig}(0,n_g)]=\P[C_{ig}]$ and $\E[D_{ig}|Z_{ig}=1,W_{ig}=w^*]-\E[D_{ig}|Z_{ig}=1,W_{ig}=w^*-1]=\P[D_{ig}(1,w^*)>D_{ig}(1,w^*-1)]=\P[GC_{ig}(w^*)]$. Finally, $\E[1-D_{ig}|Z_{ig}=1,W_{ig}=n_g]=\P[D_{ig}(1,n_g)=0]=\P[NT_{ig}]$. $\square$


\subsection{Proof of Proposition \ref{prop:lapo_ipt1}}

Under the assumptions in the proposition,
\begin{align*}
\E[Y_{ig}|D_{ig}=d,S_{ig}=s,Z_{ig}=z,W_{ig}=w]&=\sum_{\mathbf{z}_g}\E[Y_{ig}|D_{ig}=d,S_{ig}=s,Z_{ig}=z,W_{ig}=w,\mathbf{Z}_{(i)g}=\mathbf{z}_g]\\
&\quad \times \P[\mathbf{Z}_{(i)g}=\mathbf{z}_g|D_{ig}=d,S_{ig}=s,Z_{ig}=z,W_{ig}=w]\\
&=\sum_{\mathbf{z}_g}\E[Y_{ig}(d,s)|D_{ig}(z,w)=d,S_{ig}(z,\mathbf{z}_g)=s,Z_{ig}=z,\mathbf{Z}_{(i)g}=\mathbf{z}_g]\\
&\quad \times \P[\mathbf{Z}_{(i)g}=\mathbf{z}_g|D_{ig}=d,S_{ig}=s,Z_{ig}=z,W_{ig}=w]\\
&=\sum_{\mathbf{z}_g}\E[Y_{ig}(d,s)|D_{ig}(z,w)=d,S_{ig}(z,\mathbf{z}_g)=s]\\
&\quad \times \P[\mathbf{Z}_{(i)g}=\mathbf{z}_g|D_{ig}=d,S_{ig}=s,Z_{ig}=z,W_{ig}=w]\\
&=\E[Y_{ig}(d,s)|D_{ig}(z,w)=d]
\end{align*}
where the second equality uses the fact that $S_{ig}$ depends on the whole vector of instruments, the third equality follows by independence and the fourth equality uses independence of peers' types. 

Now, note that for any $s$, $\E[Y_{ig}|D_{ig}=1,S_{ig}=s,Z_{ig}=0,W_{ig}=0]=\E[Y_{ig}(1,s)|D_{ig}(0,0)=1]=\E[Y_{ig}(1,s)|AT_{ig}]$ which shows that $\E[Y_{ig}(1,s)|AT_{ig}]$ is identified. Then,
\begin{align*}
\E[Y_{ig}|D_{ig}=1,S_{ig}=s,Z_{ig}=0,W_{ig}=1]&=\E[Y_{ig}(1,s)|D_{ig}(0,1)=1]\\
&=\E[Y_{ig}(1,s)|AT_{ig}]\frac{\P[AT_{ig}]}{\P[AT_{ig}]+\P[SC(1)_{ig}]}\\
&+\E[Y_{ig}(1,s)|SC_{ig}(1)]\frac{\P[SC_{ig}(1)]}{\P[AT_{ig}]+\P[SC_{ig}(1)]}
\end{align*}
and hence $\E[Y_{ig}(1,s)|SC_{ig}(1)]$ is identified by the results above and Proposition \ref{prop:distr_comp_g}. By the same logic, 
\begin{align*}
\E[Y_{ig}|D_{ig}=1,S_{ig}=s,Z_{ig}=0,W_{ig}=2]&=\E[Y_{ig}(1,s)|D_{ig}(0,2)=1]\\
&=\E[Y_{ig}(1,s)|AT_{ig}]\frac{\P[AT_{ig}]}{\P[AT_{ig}]+\P[SC(1)_{ig}]+\P[SC(2)_{ig}]}\\
&+\E[Y_{ig}(1,s)|SC_{ig}(1)]\frac{\P[SC_{ig}(1)]}{\P[AT_{ig}]+\P[SC(1)_{ig}]+\P[SC(2)_{ig}]}\\
&+\E[Y_{ig}(1,s)|SC_{ig}(2)]\frac{\P[SC_{ig}(2)]}{\P[AT_{ig}]+\P[SC(1)_{ig}]+\P[SC(2)_{ig}]}
\end{align*}
and thus $\E[Y_{ig}(1,s)|SC_{ig}(2)]$ is identified. The same reasoning shows that as long as all required probabilities are non-zero, $\E[Y_{ig}(1,s)|\xi_{ig}]$ is identified for all values of $\xi_{ig}$ except for $\xi_{ig}=NT$, since the assignment $(1,s)$ is never observed for never-takers. Identification of $\E[Y_{ig}(0,s)|\xi_{ig}]$ for all values of $\xi_{ig}\ne AT$ follows similarly. $\square$


\section{Proofs of Additional Results}

\subsection{Proof of Lemma \ref{lemma:indirect_ITT}}

Using that:
\begin{align*}
\E[Y_{ig}|Z_{ig}=z,Z_{jg}=z']&=\E[Y_{ig}(0,0)]\\
&+\E[(Y_{ig}(1,0)-Y_{ig}(0,0))D_{ig}(z,z')(1-D_{jg}(z',z))]\\
&+\E[(Y_{ig}(0,1)-Y_{ig}(0,0))(1-D_{ig}(z,z'))D_{jg}(z',z)]\\
&+\E[(Y_{ig}(1,1)-Y_{ig}(0,0))D_{ig}(z,z')D_{jg}(z',z)],
\end{align*}
we have:
\begin{align*}
\E[Y_{ig}|Z_{ig}=1,Z_{jg}&=0]-\E[Y_{ig}|Z_{ig}=0,Z_{jg}=0]=\\
&+\E[(Y_{ig}(1,0)-Y_{ig}(0,0))(D_{ig}(0,1)(1-D_{jg}(1,0))-D_{ig}(0,0)(1-D_{jg}(0,0)))]\\
&+\E[(Y_{ig}(0,1)-Y_{ig}(0,0))((1-D_{ig}(0,1))D_{jg}(1,0)-(1-D_{ig}(0,0))D_{jg}(0,0))]\\
&+\E[(Y_{ig}(1,1)-Y_{ig}(0,0))(D_{ig}(0,1)D_{jg}(1,0)-D_{ig}(0,0)D_{jg}(0,0))].
\end{align*}

Tables \ref{tab:indirectITT1}, \ref{tab:indirectITT2} and \ref{tab:indirectITT3} list the possible values that the terms that depend on the potential treatment statuses can take, which gives the desired result after some algebra. $\square$

\begin{table}
\caption{$D_{ig}(0,1)(1-D_{jg}(1,0))-D_{ig}(0,0)(1-D_{jg}(0,0))$}
\begin{center}
\begin{tabular}{cccc|c|cc}\label{tab:indirectITT1}
$D_{ig}(1,0)$ & $D_{jg}(0,1)$ & $D_{ig}(0,0)$ & $D_{jg}(0,0)$ & Difference & $\xi_{ig}$  & $\xi_{jg}$\\ \hline \hline
1 & 1 & 1 & 1 & 0  \\
1 & 1 & 1 & 0 & -1 & AT & SC,C \\
1 & 0 & 1 & 0 & 0 \\
1 & 1 & 0 & 1 & 0 \\
1 & 1 & 0 & 0 & 0 \\
1 & 0 & 0 & 0 & 1 & SC & GC,NT \\
0 & 1 & 0 & 1 & 0 \\
0 & 1 & 0 & 0 & 0 \\
0 & 0 & 0 & 0 & 0 \\
\end{tabular}
\end{center}
\end{table}

\begin{table}
\caption{$(1-D_{ig}(0,1))D_{jg}(1,0)-(1-D_{ig}(0,0))D_{jg}(0,0)$}
\begin{center}
\begin{tabular}{cccc|c|cc}\label{tab:indirectITT2}
$D_{ig}(1,0)$ & $D_{jg}(0,1)$ & $D_{ig}(0,0)$ & $D_{jg}(0,0)$ & Difference & $\xi_{ig}$  & $\xi_{jg}$\\ \hline \hline
1 & 1 & 1 & 1 & 0 \\
1 & 1 & 1 & 0 & 0 \\
1 & 0 & 1 & 0 & 0 \\
1 & 1 & 0 & 1 & -1 & SC & AT \\
1 & 1 & 0 & 0 & 0 \\
1 & 0 & 0 & 0 & 0 \\
0 & 1 & 0 & 1 & 0 \\
0 & 1 & 0 & 0 & 1 & C,CG,NT & SC,C\\
0 & 0 & 0 & 0 & 0 \\
\end{tabular}
\end{center}
\end{table}

\begin{table}
\caption{$D_{ig}(0,1)D_{jg}(1,0)-D_{ig}(0,0)D_{jg}(0,0)$}
\begin{center}
\begin{tabular}{cccc|c|cc}\label{tab:indirectITT3}
$D_{ig}(1,0)$ & $D_{jg}(0,1)$ & $D_{ig}(0,0)$ & $D_{jg}(0,0)$ & Difference & $\xi_{ig}$  & $\xi_{jg}$\\ \hline \hline
1 & 1 & 1 & 1 & 0 \\
1 & 1 & 1 & 0 & 1 & AT & SC,C \\
1 & 0 & 1 & 0 & 0\\
1 & 1 & 0 & 1 & 1 & SC & AT \\
1 & 1 & 0 & 0 & 1 & SC & SC,C \\
1 & 0 & 0 & 0 & 0 \\
0 & 1 & 0 & 1 & 0 \\
0 & 1 & 0 & 0 & 0 \\
0 & 0 & 0 & 0 & 0 \\
\end{tabular}
\end{center}
\end{table}


\subsection{Proof of Lemma \ref{lemma:total_ITT}}

Using that:
\begin{align*}
\E[Y_{ig}|Z_{ig}=z,Z_{jg}=z']&=\E[Y_{ig}(0,0)]\\
&+\E[(Y_{ig}(1,0)-Y_{ig}(0,0))D_{ig}(z,z')(1-D_{jg}(z',z))]\\
&+\E[(Y_{ig}(0,1)-Y_{ig}(0,0))(1-D_{ig}(z,z'))D_{jg}(z',z)]\\
&+\E[(Y_{ig}(1,1)-Y_{ig}(0,0))D_{ig}(z,z')D_{jg}(z',z)],
\end{align*}
we have:
\begin{align*}
\E[Y_{ig}|Z_{ig}=1,Z_{jg}&=1]-\E[Y_{ig}|Z_{ig}=0,Z_{jg}=0]=\\
&+\E[(Y_{ig}(1,0)-Y_{ig}(0,0))(D_{ig}(1,1)(1-D_{jg}(1,1))-D_{ig}(0,0)(1-D_{jg}(0,0)))]\\
&+\E[(Y_{ig}(0,1)-Y_{ig}(0,0))((1-D_{ig}(1,1))D_{jg}(1,1)-(1-D_{ig}(0,0))D_{jg}(0,0))]\\
&+\E[(Y_{ig}(1,1)-Y_{ig}(0,0))(D_{ig}(1,1)D_{jg}(1,1)-D_{ig}(0,0)D_{jg}(0,0))].
\end{align*}

Tables \ref{tab:totalITT1}, \ref{tab:totalITT2} and \ref{tab:totalITT3} list the possible values that the terms that depend on the potential treatment statuses can take, which gives the desired result after some algebra. $\square$

\begin{table}
\caption{$D_{ig}(1,1)(1-D_{jg}(1,1))-D_{ig}(0,0)(1-D_{jg}(0,0))$}
\begin{center}
\begin{tabular}{cccc|c|cc}\label{tab:totalITT1}
$D_{ig}(1,1)$ & $D_{jg}(1,1)$ & $D_{ig}(0,0)$ & $D_{jg}(0,0)$ & Difference & $\xi_{ig}$  & $\xi_{jg}$\\ \hline \hline
1 & 1 & 1 & 1 & 0 \\
1 & 1 & 0 & 1 & 0 \\
0 & 1 & 0 & 1 & 0 \\
1 & 1 & 1 & 0 & -1 & AT & SC,C,GC \\
1 & 1 & 0 & 0 & 0 \\
0 & 1 & 0 & 0 & 0 \\
1 & 0 & 1 & 0 & 0 \\
1 & 0 & 0 & 0 & -1 & SC,C,GC & NT \\
0 & 0 & 0 & 0 & 0 \\
\end{tabular}
\end{center}
\end{table}

\begin{table}
\caption{$(1-D_{ig}(1,1))D_{jg}(1,1)-(1-D_{ig}(0,0))D_{jg}(0,0)$}
\begin{center}
\begin{tabular}{cccc|c|cc}\label{tab:totalITT2}
$D_{ig}(1,1)$ & $D_{jg}(1,1)$ & $D_{ig}(0,0)$ & $D_{jg}(0,0)$ & Difference & $\xi_{ig}$  & $\xi_{jg}$\\ \hline \hline
1 & 1 & 1 & 1 & 0 \\
1 & 1 & 0 & 1 & -1 & SC,C,GC & AT \\
0 & 1 & 0 & 1 & 0 \\
1 & 1 & 1 & 0 & 0 \\
1 & 1 & 0 & 0 & 0 \\
0 & 1 & 0 & 0 & 1 & NT & SC,C,GC \\
1 & 0 & 1 & 0 & 0 \\
1 & 0 & 0 & 0 & 0 \\
0 & 0 & 0 & 0 & 0 \\
\end{tabular}
\end{center}
\end{table}

\begin{table}
\caption{$D_{ig}(1,1)D_{jg}(1,1)-D_{ig}(0,0)D_{jg}(0,0)$}
\begin{center}
\begin{tabular}{cccc|c|cc}\label{tab:totalITT3}
$D_{ig}(1,1)$ & $D_{jg}(1,1)$ & $D_{ig}(0,0)$ & $D_{jg}(0,0)$ & Difference & $\xi_{ig}$  & $\xi_{jg}$\\ \hline \hline
1 & 1 & 1 & 1 & 0 \\
1 & 1 & 1 & 0 & 1 & SC,C,GC & AT \\
1 & 0 & 1 & 0 & 0\\
1 & 1 & 0 & 1 & 1 & AT & SC,C,GC \\
1 & 1 & 0 & 0 & 1 & SC,C,GC & SC,C,GC \\
1 & 0 & 0 & 0 & 0 \\
0 & 1 & 0 & 1 & 0 \\
0 & 1 & 0 & 0 & 0 \\
0 & 0 & 0 & 0 & 0 \\
\end{tabular}
\end{center}
\end{table}


\subsection{Proof of Proposition \ref{prop:multi}}

Under the conditions of the proposition, 
\begin{align*}
\E[\1(D_{ig}=k)|Z_{ig}=k,Z_{jg}=0]&=\E[\1(D_{ig}(k,0)=d)|Z_{ig}=k,Z_{jg}=0]=\E[\1(D_{ig}(k,0)=k)]
\end{align*}

On the other hand, for any $k\in\{0,1,\ldots,K\}$,
\begin{align*}
\E[Y_{ig}\1(D_{ig}=k)|Z_{ig}=k,Z_{jg}=0]&=\E[Y_{ig}(k,0)\1(D_{ig}(k,0)=k)]\\
\E[Y_{ig}\1(D_{ig}=0)|Z_{ig}=k,Z_{jg}=0]&=\E[Y_{ig}(0,0)\1(D_{ig}(k,0)=0)]\\
&=\E[Y_{ig}(0,0)|D_{ig}(k,0)=0]\E[\1(D_{ig}(k,0)=0)]\\
&=\E[Y_{ig}(0,0)\1(D_{ig}(k,0)=k)]+\E[Y_{ig}(0,0)\1(D_{ig}(k,0)=0)]
\end{align*}
and $\E[Y_{ig}|Z_{ig}=k,Z_{jg}=0]=\E[Y_{ig}\1(D_{ig}=k)|Z_{ig}=k,Z_{jg}=0]+\E[Y_{ig}\1(D_{ig}=0)|Z_{ig}=k,Z_{jg}=0]$, from which:
\begin{align*}
\frac{\E[Y_{ig}|Z_{ig}=k,Z_{jg}=0]-\E[Y_{ig}|Z_{ig}=0,Z_{jg}=0]}{\E[\1(D_{ig}=k)|Z_{ig}=k,Z_{jg}=0]}=\E[Y_{ig}(k,0)|D_{ig}(k,0)=k]
\end{align*}
Similarly,
\begin{align*}
\frac{\E[Y_{ig}|Z_{ig}=0,Z_{jg}=k]-\E[Y_{ig}|Z_{ig}=0,Z_{jg}=0]}{\E[\1(D_{jg}=k)|Z_{ig}=0,Z_{jg}=k]}=\E[Y_{ig}(0,k)|D_{jg}(k,0)=k]
\end{align*}
as required. $\square$


\end{appendices}

\end{document}